\documentclass[reprint,superscriptaddress,amsmath,amssymb,aps,pra]{revtex4-2}
\usepackage{graphicx}
\graphicspath{{Figures/}{figures/}}
\usepackage{dcolumn}
\usepackage{bm}
\usepackage{xcolor}
\usepackage{hyperref}

\newcommand{\LH}{\mathcal{L}_H}
\newcommand{\LHS}{\mathcal{L}_{H_S}}
\newcommand{\LHO}{\mathcal{L}_{H_0}}
\newcommand{\LHI}{\mathcal{L}_{H'}}
\newcommand{\Om}{\omega}
\newcommand{\om}{\omega_0}
\newcommand{\acc}{\bar{a}}
\newcommand{\Sa}{A}

\begin{document}
\title{General Master Equation for Two Accelerated Detectors: \\ Entanglement Generation in Common Vacuum}
\author{Riccardo Acquaviva}
\email{riccardo.acquaviva@uniba.it}
\affiliation{Dipartimento di Fisica, Universit\`a di Bari, I-70126 Bari, Italy}
\affiliation{INFN, Sezione di Bari, I-70126 Bari, Italy}
\author{Paolo Facchi}
\email{paolo.facchi@ba.infn.it}
\affiliation{Dipartimento di Fisica, Universit\`a di Bari, I-70126 Bari, Italy}
\affiliation{INFN, Sezione di Bari, I-70126 Bari, Italy}

\begin{abstract}
We derive a general master equation for two uniformly accelerated two-level detectors with different internal orientations, following the same worldline and weakly coupled to independent bosonic fields in the Minkowski vacuum. We consider coherent and dissipative contributions and allow arbitrary field masses, spacetime dimensions, and  coupling tensors. For equal orientations, we prove that singlet-triplet coherences decay exponentially when the Kossakowski matrix has rank at least two. 
For fixed couplings and a common field spectrum, we classify the acceleration dependence of stationary concurrence in terms of the  transverse coupling ratio, including acceleration-independent behavior and entanglement loss and recovery. We illustrate the framework with massless scalar fields in four spacetime dimensions, considering isotropic and mixed coupling for equal orientations. For isotropic coupling, any nonzero misalignment of the detectors yields a unique separable stationary state, while near alignment entanglement generated from a separable initial state can persist over timescales proportional to the inverse square of the misalignment angle.

\end{abstract}

\maketitle

\section{\label{sec1}INTRODUCTION}
The Unruh effect is a fundamental prediction of relativistic quantum field theory. A detector undergoing uniform proper acceleration through the Minkowski vacuum generally exhibits a nonvanishing excitation probability. More generally, an $n$-level quantum system coupled to one or more quantum fields and following a uniformly accelerated trajectory may undergo transitions to excited states even when the fields are initially  in the Minkowski vacuum. 
For constant local couplings whose bath correlations satisfy the Kubo-Martin-Schwinger (KMS) condition, the transition rates
satisfy a thermal detailed-balance relation at the Unruh temperature $T_U=a/(2\pi)$, which is determined by the acceleration~\cite{Unruh76,Davies75,Davies82,Unruh84,Takagi86,Wald94,Crispino08,Kubo57,Martin59}.

The Unruh radiation is closely related to the Hawking radiation emitted by black holes~\cite{Hawking74,Hawking75,Witten25,Mukhanov07}. 
The equivalence principle suggests a natural connection between the two effects, linking acceleration in flat spacetime to gravity~\cite{MisnerThorneWheeler1973}.  A fuller understanding of this connection also requires consideration of the quantum state of the field and the role of horizons: both the Rindler horizon~\cite{Rindler66} and the black hole event horizon restrict the regions of spacetime accessible to the respective observers~\cite{Davies82,Wald94}.

The Unruh effect can be investigated using a Unruh-DeWitt detector~\cite{Unruh76,DeWitt79}, modeled as a finite-level quantum system following a prescribed worldline and coupled locally to relativistic quantum fields  in the Minkowski vacuum. Within this framework, the transition rates between the detector energy levels can be computed directly from the field correlation functions evaluated along the detector trajectory~\cite{Davies82}. Over the years, investigations have extended far beyond the simplest models involving massless scalar fields in $3+1$ or $1+1$ spacetime dimensions. The detector response has been shown to depend on the spacetime dimensionality, the field content, and the specific detector-field coupling~\cite{Takagi86}. 

Although the Unruh effect is firmly established theoretically, its direct experimental observation remains extremely challenging, since appreciable Unruh temperatures require very large accelerations~\cite{Crispino08}. Nevertheless, several approaches to probing Unruh-related phenomena have been proposed over the past decades, including rapidly varying plasma configurations~\cite{Yablonovitch89}, ultra-intense laser configurations~\cite{Chen99,Schutzhold06}, high-$Q$ cavities~\cite{Scully03}, and superconducting quantum circuits~\cite{Felicetti15}. More recent proposals have also considered circular motion in Josephson junctions~\cite{Katayama25} and rotating molecules~\cite{ZhouHuYu25}. Related investigations in quantum information have addressed entanglement in non-inertial frames~\cite{Mann05}, as well as timelike vacuum entanglement and geometric-phase signatures of the timelike Unruh effect~\cite{Olson11,Quach22}.
An important aspect of accelerated quantum systems is the interplay between acceleration-induced effects and quantum entanglement~\cite{ZhouHuYu22, YanZhang22, Xu23, HuYu15, YangHuYu16}.

Uniformly accelerated detectors suitably coupled to quantum fields  in the Minkowski vacuum exhibit transition rates, measured with respect to their proper time, as if they were interacting with a thermal common bath at the Unruh temperature. This perspective makes it possible to go beyond the computation of excitation rates and to investigate the reduced dynamics of these accelerated detectors~\cite{Braun02, Benatti03sec, Benatti04}.  

Within this framework, this work focuses on a quantum system consisting of two uniformly accelerated two-level atoms following the same worldline and weakly coupled to a common environment composed of  bosonic scalar fields in the Minkowski vacuum~\cite{Benatti04}. 
Our formulation allows arbitrary coupling tensors, spacetime dimension, and any finite number of independent free scalar fields with arbitrary masses. Moreover, both the dissipative and effective Hamiltonian contributions have been considered and the detectors could have  arbitrary internal orientations.

By tracing out the field degrees of freedom through a Nakajima-Zwanzig projection-operator formalism~\cite{Nakajima58,Zwanzig60}, one obtains an exact equation for the reduced dynamics, which is generally non-Markovian. Under the stated weak-coupling assumption,
the reduced dynamics is described by a completely positive,
trace-preserving quantum dynamical semigroup~\cite{Gorini78,Spohn80,Alicki87,Breuer02},
obtained through the Davies weak-coupling limit~\cite{BDavies74,BDavi76sec,Facchi17}.
Its generator has the Gorini-Kossakowski-Lindblad-Sudarshan form~\cite{Gorini76,Lindblad76}.

We then investigate the entanglement properties of the stationary states through their concurrence~\cite{Wootters97,Wootters98,Wootters01}.
For equal detector orientations, we obtain the stationary state and its concurrence for a general collective coupling, before specializing to two representative choices.

The first example is an isotropic coupling, whereas the second example involves a mixed coupling tensor with a nonzero imaginary antisymmetric part.

Beyond these examples, we establish the sufficient condition on the Kossakowski matrix $\operatorname{rank}\Gamma\geq2$ for the exponential decay of all singlet-triplet coherences at equal orientations, with the effective Hamiltonian retained.

We classify the acceleration dependence in terms of the transverse coupling ratio.
This classification includes acceleration-independent entanglement, entanglement loss, and recovery at larger acceleration.

Finally, for isotropic coupling, we show that any nonzero misalignment removes the stationary entanglement permitted by exact alignment, while sufficiently small misalignments give rise to long-lived transient entanglement generated from separable initial states.

The structure of the paper is the following.
In Section~\ref{sec:2}, the quantum system under consideration is introduced and the corresponding master equation is derived.
Section~\ref{sec:equal-orientation-entanglement} determines the stationary states and their concurrence for equal orientations and classifies the acceleration dependence before analyzing the isotropic and mixed coupling examples.
Section~\ref{sec:unequal-orientations} examines the loss of stationary entanglement and the persistence of transient entanglement for unequal orientations.
The results and their implications are discussed in Section~\ref{sec:discussion}.
The appendices provide the technical details supporting the derivations.

\section{\label{sec:2}THE MASTER EQUATION}
In this section we derive a master equation describing the Markovian reduced dynamics of two uniformly accelerated atoms (detectors), weakly coupled to  scalar fields in the Minkowski vacuum.

\subsection{System}
\label{sec:system}

We consider two two-level detectors.
The free dynamics of this system in $\mathbb{C}^2\otimes\mathbb{C}^2$ is described by the Hamiltonian
\begin{equation}
H_S=\frac{\om}{2}
\sum_{a=1}^{2}
\bm n^{(a)}\cdot\bm\sigma^{(a)}.
\label{eq:system_hamiltonian}
\end{equation}
Here $\bm\sigma^{(a)}=(\sigma_1^{(a)},\sigma_2^{(a)},\sigma_3^{(a)})$, for $a=1,2$, with
\begin{equation}
\begin{split}
\sigma_i^{(1)}=\sigma_i\otimes\sigma_0,\qquad
\sigma_i^{(2)}=\sigma_0\otimes\sigma_i,
\quad i=1,2,3,
\end{split}
\label{eq:embedded-pauli}
\end{equation}
 where $\sigma_i$ are the Pauli matrices, $\sigma_0$ is the $2\times 2$ identity matrix, and
$\bm n^{(a)}=(n_1^{(a)},n_2^{(a)},n_3^{(a)})$, with $a=1,2$, are unit vectors associated with the two detectors, which specify internal energy axes relative to the interaction channels.
The parameter $\om>0$ denotes the common energy gap of the two detectors.
Throughout, repeated Cartesian indices are summed from $1$ to $3$ unless sums are displayed explicitly, and we use natural units, $\hbar=c=k_B=1$.

The dynamics of the system is described in a 
reference frame comoving with the detectors, where both detectors are at rest and their common proper time $\tau$ is used as the evolution parameter. In an inertial coordinate system of an $n$-dimensional Minkowski spacetime, their motion is described by a hyperbolic trajectory with constant proper acceleration. Both detectors are assumed to be pointlike and to follow the same uniformly accelerated worldline, parametrized as~\cite{Rindler66}
\begin{equation}
\begin{split}
x^0(\tau)&=\frac{1}{a}\sinh(a\tau),
\\
x^1(\tau)&=\frac{1}{a}\cosh(a\tau),
\\
x^i(\tau)&=0,\, i=2,\ldots,n-1,
\end{split}
\label{eq:rindler_coordinates}
\end{equation}
where $a$ denotes the constant proper acceleration. The acceleration is directed along the $x^1$ axis, while the transverse coordinates remain constant and can be set to zero without loss of generality by an appropriate choice of the spatial origin.

The detectors are weakly coupled to a common environment consisting of scalar bosonic fields. The interaction is linear in the detector observables and is written at the reference event $x(0)$ on their common worldline. Free field evolution with respect to proper time transports the field operators along $x(\tau)$, as specified below. At $\tau=0$, the interaction reads
\begin{equation}
H'=
\sum_{k=1}^{3}
S_k
\otimes\Phi_k\bigl(x(0)\bigr),
\qquad 
S_k = \sigma_k^{(1)}+\sigma_k^{(2)}.
\label{eq:H_int}
\end{equation}
The index $k$ labels the three coupling channels.
The operators $\Phi_k(x)$ describe the field degrees of freedom interacting with the detectors. No specific spacetime dimensionality is assumed. 

We consider $N$ independent real, free scalar fields, which may have different masses or be massless. 
The bath operators entering the interaction Hamiltonian can then be expressed as
\begin{equation}
\Phi_k(x)
=
\sum_{c=1}^{N}
\Big[
(\chi_k^c)^*\phi_c^{(+)}(x)
+
\chi_k^c \phi_c^{(-)}(x)
\Big],
\label{eq:field_operator}
\end{equation}
where the complex coefficients $\chi_k^c$  characterize the coupling between the $c$-th scalar field and the $k$-th  detector coupling channel and their magnitudes determine the relative coupling strengths~\cite{Benatti04}. Since
\begin{equation}
\phi_c^{(+)}(x)=
\bigl(\phi_c
^{(-)}(x)\bigr)^{\dagger} \,,
\end{equation}
each operator $\Phi_k$ is Hermitian. In Eq.~\eqref{eq:field_operator}, $\phi_c^{(+)}$ denotes the annihilation part of the field in the Minkowski mode decomposition, and $\phi_c^{(-)}=(\phi_c^{(+)})^\dagger$ denotes its creation part. The full real scalar field is then
\begin{equation}
\phi_c(x)=\phi_c^{(+)}(x)+\phi_c^{(-)}(x)\,.
\label{eq:full-scalar-field}
\end{equation}
Writing $|0_M\rangle$ for the Minkowski vacuum, we have $\phi_c^{(+)}|0_M\rangle=0$. The coefficients $\chi_k^c$ are the same for both detectors, which therefore couple collectively to the same bath operators. 

For general complex coefficients, the resulting bath observables need not be local with respect to the underlying real scalar fields~\cite{Crispino08, Funai19}. We  regard the interaction as a specified effective detector-field coupling; thermal detailed balance is not assumed for arbitrary coupling tensors.

In the comoving  frame, the field Hamiltonian $H_B$ generates the field free evolution with respect to the detectors' common proper time $\tau$. For a uniformly accelerated trajectory, with proper acceleration $a$, this time evolution is  realized by Lorentz boosts with generator $K$ and rapidity $a\tau$~\cite{Takagi86}. Thus, $H_B = a K$ and
\begin{equation}
\mathrm{e}^{i\tau H_{B}}\,
\phi_c\big(x(0)\big)\,
\mathrm{e}^{-i\tau H_{B}}=
\phi_c\big(\Lambda_{a\tau} x(0)\big) =
\phi_c\bigl(x(\tau)\bigr)\, .
\label{eq:boost}
\end{equation}
For boost rapidity \(\epsilon\), we choose the generator \(K\) so that
\(\mathrm{e}^{i\epsilon K}\phi_c(x)\mathrm{e}^{-i\epsilon K}
=\phi_c(\Lambda_\epsilon x)\), where
\(\Lambda_{a\tau}x(0)=x(\tau)\).

Denoting the coupling constant by $\lambda$, the total Hamiltonian $H$ reads
\begin{equation}
H=H_{S}+H_{B}+\lambda\,H^{\prime} \,.
\label{eq:total-hamiltonian}
\end{equation}
The density matrix $\rho_{\mathrm{tot}}$ of the total system (detectors + fields) evolves unitarily according to the von Neumann equation
\begin{equation}
\dot{\rho}_{\mathrm{tot}}(\tau)= \LH\left[\rho_{\mathrm{tot}}(\tau)\right]=-i \left[H,\rho_{\mathrm{tot}}(\tau)\right] \,,
\label{eq:von_neumann}
\end{equation}
where $\mathcal{L}_H=-i[H,{\cdot}]$ is the Liouvillian associated with~$H$.
Our main object is the reduced density matrix of the detectors 
\begin{equation}
\rho(\tau)=\operatorname{Tr}_B\bigl(\rho_{\mathrm{tot}}(\tau)\bigr). 
\end{equation}
The reduced density matrix is Hermitian ($\rho^\dagger=\rho$), has unit trace ($\operatorname{Tr}[\rho]=1$), and is positive semidefinite ($\rho\geqslant0$).

 We assume a detectors-fields factorized initial state at $\tau=0$, 
\begin{equation}
   \rho_{\mathrm{tot}}(0)=\rho(0)\otimes w_M , 
\end{equation}
with the fields being in the vacuum state $w_M=|0_M\rangle\langle0_M|$.
Since $K |0_M\rangle = 0$, the field vacuum is stationary with respect to the field free  evolution $H_B= a K$,
\begin{equation}
\mathcal{L}_{H_B} [w_M] =-i [H_B,w_M]=0\,.
\label{eq:bath-stationarity-centering1}
\end{equation}
Moreover, being linear in the fields, the bath operators have vanishing vacuum expectation value 
\begin{equation}
\bigl\langle\Phi_r(x)\bigr\rangle_M = \operatorname{Tr}_B\bigl(\Phi_r(x)\, w_M\bigr)=0\,.
\label{eq:bath-stationarity-centering2}
\end{equation}
This identity eliminates the first-order contribution in $\lambda$ to the reduced generator.

The detectors' reduced dynamics can be obtained by using the Nakajima-Zwanzig projection-operator formalism~\cite{Nakajima58, Zwanzig60}, which yields an exact integro-differential equation with a memory kernel. Retaining the leading nonvanishing contribution to this kernel in the weak-coupling expansion, the equation takes the form
\begin{equation}
\dot{\rho}(\tau)
=\LHS\!\big[\rho(\tau)\big]
+ \lambda^2 \int_{0}^{\tau}\, \mathrm{d}s\,\mathcal{K}_{\tau,s}\left[\rho(s)\right] \,,
\end{equation}
where $\LHS[\rho]=-i[H_S,\rho]$. 
With $H_0=H_S+H_B$, the memory superoperator to second order in the coupling constant is
\begin{equation}    \mathcal{K}_{\tau,s}\left[\rho(s)\right]=
\operatorname{Tr}_{B}\Big(\LHI\mathrm{e}^{(\tau-s) 
\LHO}\LHI [\rho(s)\otimes w_M]\Big).
\end{equation}

In this context, the weak-coupling limit plays a crucial role, since it provides a controlled limit in which the reduced dynamics can be described by a Markovian master equation~\cite{Spohn80,Alicki87,Facchi01,Facchi17,BDavies74,BDavi76sec}. 
In  van Hove's weak-coupling limit, the system-bath coupling constant $\lambda$ tends to zero while the rescaled macroscopic time $\tau'=\lambda^{2}\tau$ is kept fixed. Accordingly, the proper time scales as $\tau=\tau'/\lambda^{2}$, and the memory dependence becomes local on the rescaled timescale in the limit. In the interaction picture, under these assumptions, the resulting reduced dynamics converges at fixed $\tau'$ to a completely positive, trace-preserving quantum dynamical semigroup whose generator has the Gorini-Kossakowski-Lindblad-Sudarshan (GKLS) form~\cite{Gorini76,Lindblad76}. 

The weak-coupling limit also provides a secular average (Davies' average) over the free system evolution. Terms connecting distinct Bohr frequencies vanish under this average, whereas all contributions within the same frequency sector remain~\cite{BDavies74,BDavi76sec,Benatti04, KaplanekTjoa23,HanEtAl2025}. 

Notice that the field operators~\eqref{eq:field_operator} are operator-valued distributions.
The weak-coupling construction is understood for regularized
bath observables whose stationary correlation functions satisfy
the decay and integrability assumptions required by the Davies
limit~\cite{BDavies74,BDavi76sec,Facchi17, KaplanekTjoa23}.
The regularization is chosen compatibly with proper-time
stationarity along the trajectory~\cite{Schlicht2004,LoukoSatz2006}.
The pointlike expressions  are understood distributionally,
and a pointlike limit is taken only when the relevant response
functions and renormalized dispersive coefficients exist.

Having used $\lambda$ to organize the weak-coupling
expansion and define the van Hove limit, we now
absorb it into the interaction coefficients
$\chi_k^c$, retaining the same notation.
Thus, all bath-induced terms implicitly contain
the factor $\lambda^2$, and $\tau$ remains the
physical proper time. The resulting master
equation in the Schr\"odinger picture reads
\begin{equation}
\dot{\rho}(\tau)=\LHS[\rho (\tau)]+ 
\mathcal{K}^{\sharp}[\rho (\tau)] \,.
\label{eq:master_equation}
\end{equation}
Here, the GKLS superoperator is
\begin{equation}
   \mathcal{K}^{\sharp}[\rho]
=-i[H_L,\rho]+\mathcal{L}[\rho], 
\label{eq:K_superoperator}
\end{equation}
where
\begin{align}
\mathcal L[\rho]
 &=
\!\!\sum_{\Om=0,\pm\om}\!\! \alpha_{jk}(\Om) \Bigl(
S_k^{(\Om)}\rho S_j^{(\Om)\dagger}-\frac12
\bigl\{S_j^{(\Om)\dagger}S_k^{(\Om)},\rho\bigr\}
\Bigr),\nonumber\\
H_L &= \frac{1}{2} 
\sum_{\Om=0,\pm\om} \!\!
\kappa_{jk}(\Om)\,
S_j^{(\Om)\dagger}S_k^{(\Om)},
\label{eq:K_superoperator2}
\end{align}
In Eq.~\eqref{eq:K_superoperator2}, $\Om=0,\pm\om$ labels the three Bohr-frequency sectors that contribute to the dynamics. Although the two-detector Hamiltonian also has Bohr frequencies $\Om=\pm2\om$, the corresponding spectral components of the collective coupling operators vanish. 

The jump operators entering Eq.~\eqref{eq:K_superoperator2} are
\begin{equation}
\begin{split}
  S_{k}^{(0)} & =\sum_{a=1,2}(P_a)_{kl}\,\sigma_{l}^{(a)},
\\
S_{k}^{(\pm\om)} &=S_{k}^{(\mp\om)\dagger} =\frac{1}{2} \sum_{a=1,2}(Q_a \pm i \Omega_a)_{kl}\,\sigma_{l}^{(a)},  
\end{split}
\label{eq:jumps1}
\end{equation}
where for each detector $a=1,2$, we introduced the real $3\times3$ matrices
\begin{equation}
\begin{split}
P_a=\bm n^{(a)}\bm n^{(a)T}\,, \quad 
Q_a=I_3-P_a \,,  \quad
(\Omega_a)_{ij}=
\varepsilon_{ijk}n_k^{(a)}\,. 
\label{eq:detector-geometric-matrices}
\end{split}
\end{equation}
Here, $\varepsilon_{ijk}$ denotes the Levi-Civita symbol with $\varepsilon_{123}=1$.
The matrices $P_a$ and $Q_a$ project onto the directions parallel and perpendicular to the detector orientation $\bm n^{(a)}$, respectively, while $\Omega_a^T=-\Omega_a$ describes rotations in the transverse plane,
$\Omega_a \bm{v}= \bm{v} \wedge \bm{n}^{(a)}$, for any $\bm{v}\in\mathbb{R}^3$. Together, they provide a convenient geometric representation of the detector operators.

These definitions imply
\begin{equation}
\LHS[S_k^{(\Om)}]=-i\Om S_k^{(\Om)},
\qquad 
\bigl(S_k^{(\Om)}\bigr)^\dagger=S_k^{(-\Om)}.
\label{eq:bohr-component-identities}
\end{equation}
Thus $\mathrm{e}^{-\tau\LHS }[S_r^{(\Om)}]=\mathrm{e}^{i\Om\tau}S_r^{(\Om)}$,
which makes the frequency selection in the Davies average explicit.

The coefficients in the GKLS superoperator~\eqref{eq:K_superoperator} are
\begin{eqnarray}
       \alpha_{rs}(\Om)
&=&\sum_{c=1}^{N} 
F_{n}^{(c)}(\Om) \, \chi_{rs}^{(c)} \,  ,
\label{eq:alpha}
\\
\kappa_{rs}(\Om)
&=&\sum_{c=1}^{N} 
 K_{n}^{(c)}(\Om)\, \chi_{rs}^{(c)} \, .
\label{eq:beta}
\end{eqnarray}
They depend on the $3\times3$ Hermitian coupling tensors
\begin{equation}
\chi_{rs}^{(c)}=(\chi_{r}^{c})^*  \chi_{s}^{c} =\chi_{rs}^{(S,c)}+i \,\chi_{rs}^{(A,c)} \,,
\label{eq:coupling_tensor}
\end{equation}
for each field $c=1,\dots,N$,
the response functions 
\begin{equation}
    F_{n}^{(c)}(\Om) = \int_{\mathbb{R}} \mathrm{d}\tau \, \mathrm{e}^{-i \Om \tau}\bigl\langle 
    \phi_{c}\bigl(x(\tau)\bigr) \phi_{c}\bigl(x(0)\bigr) 
    \bigr\rangle_M,
    \label{eq:responsefunc}
\end{equation}
and their Hilbert transforms
\begin{equation}
    K_{n}^{(c)}(\Om) = \frac{\mathcal{P}}{\pi} \int_{\mathbb{R}} \mathrm{d} z \frac{F_{n}^{(c)}(z)}{z-\Om}\, .
    \label{eq:Hilberttransf}
\end{equation}
Here the Hermitian tensors are written in terms of a real symmetric tensor $\chi^{(S,c)}$ and a real antisymmetric tensor $\chi^{(A,c)}$, and $\mathcal P$ denotes the Cauchy principal value at the pole. This prescription does not remove possible ultraviolet divergences at large $z$; the dispersive functions are understood with a specified ultraviolet regularization and, where required, subtraction. Appendix~\ref{app:lamb_shift} gives the subtraction used in the four-dimensional massless examples.
Both functions depend on the
vacuum Wightman two-point functions of the fields
evaluated along the uniformly accelerated worldline of the detectors in an $n$-dimensional Minkowski spacetime~\cite{Wightman56,Takagi86,Davies82}. Although the detector dynamics is formulated in the comoving frame, the field correlation functions can be evaluated using inertial Minkowski coordinates. 

Notice that both $F_n^{(c)}(\Om)$ and $K_n^{(c)}(\Om)$ are real-valued functions, so that the tensors $\alpha$ and $\kappa$ are both Hermitian. 
Moreover, since the Minkowski vacuum restricted to the Rindler wedge is a KMS
state with respect to boosts, each response function satisfies the KMS
detailed-balance relation~\cite{Bisognano75,Sewell82,Takagi86}
\begin{equation}
F_n^{(c)}(\Om)=\mathrm{e}^{-\beta\Om}F_n^{(c)}(-\Om),
\qquad \beta=\frac{2\pi}{a},
\label{eq:thermal-detailed-balance}
\end{equation}
where $\beta$ is the inverse Unruh temperature
associated with the proper acceleration
$a$~\cite{Unruh76}.
The scalar KMS relation~\eqref{eq:thermal-detailed-balance}
does not necessarily imply thermal detailed balance for the
effective bath operators $\Phi_j$. Indeed,
Eqs.~\eqref{eq:alpha} and~\eqref{eq:coupling_tensor} give
\begin{equation}
\alpha_{jk}(\Om)
-\mathrm{e}^{-\beta\Om}\alpha_{kj}(-\Om) =
2i\sum_{c=1}^N F_n^{(c)}(\Om)
\chi_{jk}^{(A,c)}.
\label{eq:effective-detailed-balance}
\end{equation}
Thus, detailed balance holds for real coupling
coefficients $\chi_k^c$, whereas complex couplings
may violate it. This is consistent with the possible
nonlocality of the effective bath observables
introduced in Eq.~\eqref{eq:field_operator}.

The decomposition~\eqref{eq:coupling_tensor} induces analogous decompositions 
\begin{equation}
\begin{split}
    \alpha (\omega) &= \alpha^{(S)}(\omega) + i \alpha^{(A)}(\omega), \\
    \kappa (\omega) &= \kappa^{(S)}(\omega) + i \kappa^{(A)}(\omega),
\end{split}
\end{equation}
of the $\alpha$ and $\kappa$ in terms of their symmetric and antisymmetric parts.

Finally, notice that the Gram factorization of the coupling tensors~\eqref{eq:coupling_tensor} gives an alternative diagonal form of the GKLS generator. For each field $c=1,\dots,N$ define the jump operators
\begin{equation}
L_{c}^{(\Om)}=
\chi_j^c \, S_j^{(\Om)},
\qquad \Om=0,\pm\om.
\label{eq:collective-jump}
\end{equation}
Then, the GKLS superoperator has the form~\eqref{eq:K_superoperator}
where
\begin{equation}
\begin{split}
H_L&=\frac{1}{2}\sum_{\substack{\Om=0,\pm\om\\c=1,\dots,N}}\!\!  {K_n^{(c)}(\Om)}\, L_{c}^{(\Om)\dagger} L_{c}^{(\Om)},
\\
\mathcal L
&=\!\!\sum_{\substack{\Om=0,\pm\om\\ c=1,\dots,N}} \!\! F_n^{(c)}(\Om)\, \mathcal{D}_{L_{c}^{(\Om)}},
\end{split}
\label{eq:jumps2}
\end{equation}
with
\begin{equation}
    \mathcal{D}_L[\rho]=L\rho L^\dagger-\tfrac12\{L^\dagger L,\rho\}
    \label{eq:DLdef}
\end{equation}
being the dissipator in GKLS form associated to the jump operator $L$.
Notice that, at variance with the expression~\eqref{eq:K_superoperator2}, here $\mathcal{L}$ has a diagonal form, with coefficients~$F_n^{(c)}(\omega)$. Being the Fourier transform~\eqref{eq:responsefunc} of the positive-type correlation function $\langle 
\phi_{c}\bigl(x(\tau)\bigr) \phi_{c}\bigl(x(0)\bigr)\rangle_M$ ,  $F_n^{(c)}(\omega)$ is nonnegative by Bochner's theorem. Thus it makes the complete positivity of the dynamical semigroup generated by $\mathcal{L}$ explicit.

The Hamiltonian and the dissipator in Eqs.~\eqref{eq:K_superoperator2}
and in~\eqref{eq:jumps2}, together with their Cartesian form that we will give in the next subsection,
constitute one of the main results of this work. These expressions give the  generator $\mathcal{K}^{\sharp}$ in~\eqref{eq:K_superoperator} explicitly in terms of the detector orientations $\bm{n}^{(a)}$, the symmetric and antisymmetric parts of the coupling tensors $\chi^{(c)}$, and the response functions $F_n^{(c)}$ and dispersive functions $K_n^{(c)}$ of each field.

\subsection{Cartesian expansions}

One can obtain a third useful expression of the superoperator $\mathcal{K}^{\sharp}$ directly in terms of the three Pauli operators $\sigma_j^{(a)}$, $j=1,2,3$ acting on the two detectors $a=1,2$, by plugging the explicit expression~\eqref{eq:jumps1} of the jump operators into Eqs.~\eqref{eq:K_superoperator2}.

The Hamiltonian and dissipative contributions in~\eqref{eq:master_equation} will then be expressed through
the following $3\times3$ blocks, for $a,b=1,2$:
\begin{align}
\Lambda_{ab} 
&= P_a \kappa_+(0) P_b
+\frac{1}{2}
\bigl(Q_a\kappa_+(\omega_0)Q_b-\Omega_a\kappa_+(\omega_0)\Omega_b\bigr)
\nonumber\\
& \phantom{ = } +\frac{i}2
\bigl(\Omega_a\kappa_-(\omega_0)Q_b+Q_a\kappa_-(\omega_0)\Omega_b\bigr),
\label{eq:coherent-blocks}
\end{align}
and
\begin{align}
\Gamma_{ab} 
&= P_a \alpha_+(0) P_b
+\frac{1}{2}
\bigl(Q_a\alpha_+(\omega_0)Q_b-\Omega_a\alpha_+(\omega_0)\Omega_b\bigr)
\nonumber\\
& \phantom{ = } +\frac{i}2
\bigl(\Omega_a\alpha_-(\omega_0)Q_b+Q_a\alpha_-(\omega_0)\Omega_b\bigr),
\label{eq:dissipative-blocks}
\end{align}
where
\begin{align}
    \alpha_{\pm}(\omega) &= \frac{1}{2} \bigl(\alpha(\omega)\pm \alpha(-\omega)\bigr) 
    \nonumber\\
    \kappa_{\pm}(\omega) &= \frac{1}{2} \bigl(\kappa(\omega)\pm \kappa(-\omega)\bigr) 
\end{align}
with $\alpha$ and $\kappa$ the coupling tensors~\eqref{eq:alpha} and~\eqref{eq:beta}. 
These blocks satisfy
$\Lambda_{ba}=\Lambda_{ab}^{\dagger}$ and
$\Gamma_{ba}=\Gamma_{ab}^{\dagger}$.

The dissipative contribution reads
\begin{equation}
\mathcal L[\rho]
=\sum_{a,b=1}^{2}
(\Gamma_{ab})_{ij}
\Bigl(
\sigma_j^{(b)}\rho\sigma_i^{(a)}
-\frac12\bigl\{\sigma_i^{(a)}\sigma_j^{(b)},\rho\bigr\}
\Bigr).
\label{eq:dissipative_term}
\end{equation}
The four blocks $\Gamma_{ab}$ form the Hermitian, positive-semidefinite
$6\times6$ Kossakowski matrix. The terms with $a=b$ and $a\ne b$ describe
the local and cross-detector dissipative contributions, respectively.

The Lamb-shift Hamiltonian is
\begin{equation}
H_L=\frac12\sum_{a,b=1}^{2} 
(\Lambda_{ab})_{ij}\,\sigma_i^{(a)}\sigma_j^{(b)}.
\label{eq:lamb-blocks}
\end{equation}
For $a\ne b$, operators belonging to different detectors commute;
combining the two ordered contributions and using
$\Lambda_{21}=\Lambda_{12}^{\dagger}$ gives 
the cross-detector interaction
\begin{equation}
H^{(12)}_L
=
\bigl(\operatorname{Re}\Lambda_{12}\bigr)_{ij}
\sigma_i^{(1)}\sigma_j^{(2)}.
\label{eq:H_12}
\end{equation}
This is the cross-detector contribution to the effective
Hamiltonian, $H_{\mathrm{eff}}^{(12)}=H_L^{(12)}$.
For $a=b$, the Pauli identity
\begin{equation}
\sigma_j^{(a)}\sigma_k^{(a)}
=\delta_{jk}I_4+i 
\varepsilon_{jkl}\sigma_l^{(a)}
\end{equation}
reduces the local contribution to 
\begin{align}
    H_L^{(a)} &= \frac{i}{2} (\Lambda_{aa})_{jk} \varepsilon_{jkl} \sigma_l^{(a)} 
    \nonumber\\
    &= - \frac{1}{2} \operatorname{Tr}\bigl(Q_a \kappa_{-}^{(S)}(\omega_0) - \Omega_a \kappa_{+}^{(A)}(\omega_0)\bigr)\,\bm{n}^{(a)} \cdot \bm{\sigma}^{(a)},
\end{align}
up to the irrelevant scalar $\frac12 \operatorname{Tr}(\Lambda_{aa})$.
Thus, the local Lamb shift renormalizes the energy gap while preserving
the detector's quantization axis.

The GKLS master equation~\eqref{eq:master_equation} can be rewritten as
\begin{equation}
 \dot{\rho}(\tau) = -i\bigl[H_{\text{eff}}, \rho(\tau)\bigr] + \mathcal{L}[\rho(\tau)],  
 \label{eq:rewritten_master_equation}
\end{equation}
with 
\begin{equation}
  H_{\text{eff}} = H_S + H_L = H_{\text{eff}}^{(1)} + H_{\text{eff}}^{(2)} + H_{\text{eff}}^{(12)}.
  \label{eq:H_eff_sum}
\end{equation}

Up to an additive scalar, the local Hamiltonian contribution is
\begin{equation}
H_{\mathrm{eff}}^{(a)}
=H_S^{(a)}+H_L^{(a)}
=\frac12\bigl(\om+\delta\omega_a\bigr)
\,\bm n^{(a)}\cdot\bm\sigma^{(a)},
\label{eq:H_a}
\end{equation}
where 
\begin{equation}
\delta\omega_a
=\operatorname{Tr}\Bigl(
\Omega_a\kappa_+^{(A)}(\om)
-Q_a\kappa_-^{(S)}(\om)
\Bigr)
\label{eq:coupling_parameters}
\end{equation}
is the local Lamb shift of the detector
energy gap.

Equation~\eqref{eq:H_a} describes the local coherent corrections, Eq.~\eqref{eq:H_12} gives the field-mediated coherent coupling, and Eq.~\eqref{eq:dissipative_term} contains the local and cross-detector dissipative terms.
Together, the Hamiltonian and dissipative contributions define a GKLS generator, so that the reduced evolution forms a completely positive, trace-preserving Markovian semigroup. The derivation and the expanded Cartesian expressions are given in
Appendix~\ref{app:derivationGKLS}.

The generality of the result refers to the model defined above: two co-located detectors with equal positive energy gaps, collective coupling to independent free scalar fields, and evolution with respect to a common proper time in the stationary weak-coupling regime. Within this setting, the spacetime dimension and masses enter through $F_{n}^{(c)}$ and $K_{n}^{(c)}$. The number of fields can be any finite $N$, and the coupling tensors are those generated by the coefficients $\chi_k^c$ in Eq.~\eqref{eq:coupling_tensor}. Their positive-semidefinite Gram structure and the positivity of the bath spectra ensure the positivity required by the spectral GKLS construction.
This construction assumes well-defined response functions at the relevant Bohr frequencies and sufficient decay of bath correlations for the weak-coupling limit. Pointlike couplings and dispersive terms require an ultraviolet prescription; the massless scalar field in $1+1$ dimensions additionally requires an infrared prescription. Dimensional and mass generality is understood subject to these conditions.

\subsection{Detectors with equal orientations}
For $\bm n^{(1)}=\bm n^{(2)}=\bm n$, the geometric matrices and
the Lamb shift in Eq.~\eqref{eq:coupling_parameters} are independent
of the detector index, which we omit. The Cartesian blocks in
Eqs.~\eqref{eq:coherent-blocks} and~\eqref{eq:dissipative-blocks}
coincide: $\Lambda_{ab}=\Lambda$ and $\Gamma_{ab}=\Gamma$.
The master equation therefore takes the collective form
\begin{equation}
\dot\rho= -i[H_{\mathrm{eff}},\rho]+\Gamma_{jk}\left(S_k\rho S_j
-\tfrac12\{S_jS_k,\rho\}\right),
\label{eq:master_equation_equalorientations}
\end{equation}
where $S_j=\sigma_j^{(1)}+\sigma_j^{(2)}$.

Choose a common  basis with $\bm n=(0,0,1)$ and define
\begin{equation}
  S_\pm=\frac12(S_1\pm iS_2).  
\end{equation}
The dissipator then reduces to three
collective channels,
\begin{equation}
\mathcal L[\rho]
=\gamma_\downarrow\mathcal D_{S_-}[\rho]
+\gamma_\uparrow\mathcal D_{S_+}[\rho]
+\gamma_0\mathcal D_{S_3}[\rho],
\label{eq:three-channels}
\end{equation}
with
\begin{equation}
\begin{split}
\gamma_\downarrow
&=\operatorname{Tr}\!\left[(Q-i\Omega)\alpha(-\om)\right],\\
\gamma_\uparrow
&=\operatorname{Tr}\!\left[(Q+i\Omega)\alpha(\om)\right],\\
\gamma_0&=\operatorname{Tr}[P\alpha(0)].
\end{split}
\label{eq:three-rates-tensors}
\end{equation}
These rates are nonnegative, and the eigenvalues of the
$3\times3$ matrix $\Gamma$ 
\begin{equation}
\Gamma=\gamma_0P
+\frac{\gamma_\downarrow+\gamma_\uparrow}{4}Q
+\frac{i(\gamma_\uparrow-\gamma_\downarrow)}{4}\Omega,
\label{eq:equal-orientation-gamma}
\end{equation}
are
$\gamma_\downarrow/2$, $\gamma_\uparrow/2$, and $\gamma_0$.

In the same basis, the common coherent block  reads
\begin{equation}
\Lambda
=J_\parallel P+J_\perp Q
-i\frac{\delta\omega}{2}\Omega,
\label{eq:equal-orientation-lambda}
\end{equation}
where
\begin{equation}
\begin{split}
J_{\parallel}
 &=\operatorname{Tr}\bigl(P\kappa_+^{(S)}(0)\bigr),\\
J_\perp
& 
=\frac12\operatorname{Tr}\bigl(
Q\kappa_+^{(S)}(\om)-\Omega\kappa_-^{(A)}(\om)
\bigr).
\end{split}
\label{eq:transverse-coherent-coupling}
\end{equation}
are the longitudinal and transverse couplings.
Substituting this expression into
$H_L=\frac12\Lambda_{jk}S_jS_k$ gives
\begin{align}
H_{\mathrm{eff}}
&=\frac{\om+\delta\omega}{2}S_3 +\frac{J_\parallel}{2}S_3^2+\frac{J_\perp}{2}(S_1^2+S_2^2)
\nonumber\\
&=\frac{\om+\delta\omega}{2}S_3+\frac{J_\parallel-J_\perp}{2}S_3^2 +\frac{J_\perp}{2}\bm{S}^2.
\label{eq:collective-hamiltonian}
\end{align}
The Hamiltonian is diagonal in the singlet-triplet basis of eigenvectors of $\bm{S}^2$ and $S_3$,
which we will use in the next section.

\section{Stationary states and entanglement for equal orientations}
\label{sec:equal-orientation-entanglement}

We first study the stationary regime for equal detector orientations, choosing the common axis as $\bm n=(0,0,1)$, and quantify the entanglement of the stationary states through their concurrence. We classify its dependence on acceleration, coupling asymmetry and initial preparation. We then examine an  isotropic coupling tensor and one with both symmetric and antisymmetric components.

We use the ordered singlet--triplet basis, comprising one antisymmetric singlet and three symmetric triplet states,
\begin{equation}
\mathcal B_{\mathrm{ST}}=(|s\rangle,|t_-\rangle,|t_0\rangle,|t_+\rangle),
\label{eq:singlet-triplet-basis}
\end{equation}
where
\begin{equation}
\begin{split}
|s\rangle &= \frac{1}{\sqrt{2}}(|eg\rangle-|ge\rangle),\\
|t_-\rangle = |gg\rangle,\quad|t_0\rangle &= \frac{1}{\sqrt{2}}(|eg\rangle+|ge\rangle),\quad
|t_+\rangle = |ee\rangle.
\end{split}
\label{eq:singlet-triplet-states}
\end{equation}
Here $|e\rangle$ and $|g\rangle$ are the eigenstates of
$\sigma_3$ with eigenvalues $+1$ and $-1$, respectively.
The advantage of this basis is that it diagonalizes the  effective Hamiltonian~\eqref{eq:collective-hamiltonian}, which reads
\begin{equation}
H_{\mathrm{eff}}
=E_{-}
|t_-\rangle\langle t_-|+ E_0|t_0\rangle\langle t_0|
+E_{+}
|t_+\rangle\langle t_+|,
\label{eq:hamiltonian-singlet-triplet}
\end{equation}
where $E_{0} = 4 J_\perp$ and
$E_{\pm} = 2 J_\parallel + 2 J_\perp \pm (\omega_0 + \delta\omega)$.

The collective coupling operators annihilate the singlet,
$S_i|s\rangle=0$, and the effective Hamiltonian preserves it.
Its population is therefore conserved:
\begin{equation}
p_s=\langle s|\rho(\tau)|s\rangle
=\langle s|\rho(0)|s\rangle,
\qquad 0\leq p_s\leq1.
\label{eq:singlet-weight}
\end{equation}

The triplet populations 
\begin{equation}
    p_m(\tau)=\langle t_m|\rho(\tau)|t_m\rangle, \qquad m\in\{-,0,+\},
\end{equation}
give the probabilities of occupying the three triplet states. 
In this basis, the collective operators read
\begin{equation}
\begin{split}
S_-&=\sqrt{2}\bigl(
|t_0\rangle\langle t_+|
+|t_-\rangle\langle t_0|
\bigr), \qquad
S_+ =S_-^\dagger,\\
S_3&=2\bigl(
|t_+\rangle\langle t_+|
-|t_-\rangle\langle t_-|
\bigr).
\end{split}
\end{equation}
Since both $H_{\mathrm{eff}}$ and $S_3$ are diagonal,
the Hamiltonian commutator and the dephasing channel $\mathcal{D}_{S_3}$
preserve the populations.
Taking the diagonal matrix elements of the master
equation, with the dissipator in
Eq.~\eqref{eq:three-channels}, therefore gives the population equations:
\begin{equation}
\begin{split}
\dot p_+&=-2\gamma_\downarrow p_++2\gamma_\uparrow p_0,\\
\dot p_0&=2\gamma_\downarrow(p_+-p_0)+2\gamma_\uparrow(p_--p_0),\\
\dot p_-&=2\gamma_\downarrow p_0-2\gamma_\uparrow p_-.
\end{split}
\label{eq:triplet-populations}
\end{equation}
The terms proportional to $\gamma_\downarrow$ describe the downward transitions $|t_+\rangle\to|t_0\rangle\to|t_-\rangle$, while those proportional to $\gamma_\uparrow$ describe the reverse transitions. Adding the population equations gives $\dot p_-+\dot p_0+\dot p_+=0$, preserving the total triplet weight $p_-+p_0+p_+=1-p_s$.

For $\gamma_\downarrow>0$, define the transition-rate ratio 
\begin{equation}
q=\frac{\gamma_\uparrow}{\gamma_\downarrow},
\label{eq:stationary-rate-ratio}
\end{equation}
The stationary population equations give $p_0=q p_-$ and
$p_+=q p_0$. Together with the conserved triplet weight,
these relations determine the stationary populations.

To establish convergence of the full state, we assume the sufficient condition 
\begin{equation}
    \operatorname{rank}(\Gamma)\geq2
    \label{eq:rank-condition}
\end{equation}
Since the eigenvalues of $\Gamma$ in~\eqref{eq:equal-orientation-gamma} are $\gamma_\downarrow/2$, $\gamma_\uparrow/2$ and $\gamma_0$, this requires at least two positive rates, including at least one transition rate. The singlet--triplet coherences then decay exponentially, as proved in Appendix~\ref{app:decay_antisymm}, while direct evaluation of the master equation gives the same conclusion for coherences within the triplet sector. Consequently, all off-diagonal elements vanish at long times and, for each value of $p_s$, the dynamics converges to the unique stationary state
\begin{align}
\hat\rho
&=p_s|s\rangle\langle s|+p_-|t_-\rangle\langle t_-|
+p_0|t_0\rangle\langle t_0|+p_+|t_+\rangle\langle t_+|
\nonumber\\
&=p_s|s\rangle\langle s|+\frac{1-p_s}{Z_{q}}
\bigl(|t_-\rangle\langle t_-|
+q|t_0\rangle\langle t_0|
+q^2|t_+\rangle\langle t_+|\bigr),
\label{eq:stationary-triplet}
\end{align}
where $Z_{q}=1+q+q^2$.
Here $\rho(\tau)\to\hat\rho$ as $\tau\to+\infty$.
This state satisfies $\mathcal L[\hat\rho]=0$ and commutes with
$H_{\mathrm{eff}}$, establishing stationarity under the full generator.
The conserved singlet population therefore parametrizes a family of
stationary states. For $q=0$, all triplet population occupies
$|t_-\rangle$. When $\gamma_\downarrow=0$ and $\gamma_\uparrow>0$,
the stationary state is obtained by taking $q\to+\infty$,
with all triplet population in $|t_+\rangle$.

We quantify stationary entanglement through the concurrence, which vanishes for separable states and reaches the value 1 for maximally entangled states~\cite{Wootters98, Wootters01, Wootters97}:
\begin{equation}
C[\rho]=\max\{0,\mu_1-\mu_2-\mu_3-\mu_4\},
\label{eq:concurrence-definition}
\end{equation}
where $\mu_1\geq\mu_2\geq\mu_3\geq\mu_4$ are the eigenvalues of 
\begin{equation}
   \mathcal{M}=\sqrt{\sqrt{\rho}\,\rho^{\mathrm{f}}\,\sqrt{\rho}}, 
   \qquad \rho^{\mathrm{f}}=(\sigma_2\otimes\sigma_2)\rho^T(\sigma_2\otimes\sigma_2).
\end{equation}
The spin-flip transformation associated with $\sigma_2\otimes\sigma_2$ leaves the singlet and central-triplet projections unchanged and exchanges the outer-triplet projections. Thus,
\begin{equation}
\hat{\rho}^{\mathrm{f}}= p_s |s\rangle\langle s | + p_+ |t_-\rangle\langle t_- | + p_0 |t_0\rangle\langle t_0 | +p_- |t_+\rangle\langle t_+ |
\label{eq:rho-tilde-matrix}
\end{equation}
and the unordered eigenvalues of $\mathcal M$ are $p_s$, $p_0$, $\sqrt{p_-p_+}$ and $\sqrt{p_-p_+}$. Substitution into Eq.~\eqref{eq:concurrence-definition} gives
\begin{equation}
C[\hat\rho]=\max\{0,|p_s-p_0|-2\sqrt{p_-p_+}\}.
\label{eq:singlet-triplet-concurrence}
\end{equation}
Since the stationary populations satisfy $\sqrt{p_-p_+}=p_0$,
entanglement requires $p_s>3p_0$, and the concurrence reduces to
\begin{equation}
C[\hat\rho]
=\max\left\{0,p_s-\frac{3q (1-p_s)}{1+q+q^2}\right\}.
\label{eq:stationary-concurrence-general}
\end{equation}
This is the general stationary concurrence for equal orientations under
the assumed convergence condition. Within the assumptions of
Sec.~\ref{sec:2}, spacetime dimension and field masses enter through
the transition rates. The initial preparation enters through $p_s$,
while the competition between decay and excitation enters through
$q$. Exchanging two positive transition rates sends
$q$ to $q^{-1}$ and exchanges $p_-$ and $p_+$,
leaving the concurrence unchanged. If one transition rate vanishes,
the stationary population $p_0$ is zero and $C[\hat\rho]=p_s$;
these cases correspond to $q=0$ and the limit
$q\to+\infty$.

\subsection{Classification by transverse coupling asymmetry}
\label{sec:classification}

Consider two detectors with equal internal orientations, following the
same uniformly accelerated trajectory in the Minkowski vacuum and
coupled to independent scalar fields of equal masses.
Since the fields share the same response function, we write
$F_n^{(c)}=F_n$ and define
\begin{equation}
\chi_{jk}=\sum_{c=1}^N\chi_{jk}^{(c)}
=\chi_{jk}^{(S)}+i\chi_{jk}^{(A)}.
\label{eq:total-coupling-tensor}
\end{equation}
Define the dimensionless acceleration
\begin{equation}
\acc=\frac{a}{\om}.
\label{eq:acc_dimensionless}
\end{equation}
For a nonzero decay coupling, define the transverse coupling ratio, or coupling asymmetry,
\begin{equation}
q_{\chi}
=\frac{\operatorname{Tr}\bigl((Q+i\Omega)\chi\bigr)}
{\operatorname{Tr}\bigl((Q-i\Omega)\chi\bigr)}.
\label{eq:asymptotic-rate-ratio}
\end{equation}
It depends only on the coupling tensor $\chi$ and the common detector
orientation and is independent of acceleration. 
Positivity of $\chi$ and $Q\pm i\Omega$ makes both traces
nonnegative, so $q_{\chi}\geq0$. 
If the denominator vanishes, the assumed rank
condition~\eqref{eq:rank-condition} requires a nonzero excitation
coupling; this case is treated as the limit $q_{\chi}\to+\infty$.

The transverse coupling ratio $q_{\chi}$ quantifies
the imbalance between the excitation and decay coupling weights.
For $q_{\chi}=1$, excitation and decay have equal coupling weights.
Values below 1 favour the decay channel, while values above 1
favour the excitation channel at the level of the coupling weights.
We study acceleration dependence at fixed couplings $q_{\chi}$ and fixed $p_s$.

Every finite value $q_{\chi}\geq0$ can be realized by admissible
couplings. For $\bm n=(0,0,1)$, one possible choice is
\begin{equation}
\chi=\begin{pmatrix}
1&-i\dfrac{1-q_{\chi}}{1+q_{\chi}}&0\\
i\dfrac{1-q_{\chi}}{1+q_{\chi}}&1&0\\
0&0&1
\end{pmatrix}.
\label{eq:phase-realization}
\end{equation}
Its eigenvalues are $2/(1+q_{\chi})$,
$2q_{\chi}/(1+q_{\chi})$ and $1$, all nonnegative.
Its Gram decomposition realizes these couplings with at most three
independent fields. The matrix has a finite limit as
$q_{\chi}\to+\infty$, which realizes the excitation-only endpoint. Since $\operatorname{Tr}\bigl((Q\pm i\Omega)\chi\bigr) =\operatorname{Tr}(Q\chi^{(S)})\mp\operatorname{Tr}(\Omega\chi^{(A)})$, one has $q_{\chi}\neq1$ only if $\chi^{(A)}$ has a component along $\bm n$. By Eq.~\eqref{eq:effective-detailed-balance}, real coefficients $\chi_k^c$, i.e.\ couplings local in the fields $\phi_c$, give $q_{\chi}=1$; the realization above is complex whenever $q_{\chi}\neq1$.

Using Eq.~\eqref{eq:three-rates-tensors} and the field
KMS relation~\eqref{eq:thermal-detailed-balance}, the transition-rate
ratio~\eqref{eq:stationary-rate-ratio} becomes
\begin{equation}
q
=q_{\chi}\frac{F_n(\om)}{F_n(-\om)}
=q_{\chi}\,\mathrm{e}^{-\beta\om}
=q_{\chi}\,\mathrm{e}^{-2\pi/\acc},
\label{eq:thermal-stationary-ratio}
\end{equation}
where $\beta=2\pi/a$ is the inverse Unruh temperature.
In particular, 
\begin{equation}
   q_{\chi}=\lim_{\acc\to+\infty}q(\acc), 
\end{equation}
which 
identifies the transverse coupling ratio $q_{\chi}$ as the
asymptotic transition-rate ratio.

At finite positive acceleration, $q_{\chi}=0$ suppresses excitation,
whereas $q_{\chi}\to+\infty$ describes zero decay
coupling and nonzero excitation. Both endpoints give
$C[\hat\rho]=p_s$, independently of acceleration. At either endpoint,
the rank condition~\eqref{eq:rank-condition} requires $\gamma_0>0$.

We now consider $0<q_{\chi}<\infty$, where both transition rates $\gamma_\downarrow$ and $\gamma_\uparrow$
are positive. At fixed coupling,
\begin{equation}
\frac{\partial q}{\partial \acc}
=\frac{2\pi}{\acc^2}q>0.
\label{eq:rate-ratio-derivative}
\end{equation}
Thus, increasing acceleration increases $q$ monotonically
from zero towards $q_{\chi}$. For nonzero triplet weight,
$q<1$ gives $p_->p_0>p_+$, while $q>1$ gives the
inverted ordering $p_+>p_0>p_-$ in~\eqref{eq:stationary-triplet}. At $q=1$, the transition
rates and the three triplet populations are equal.
Since
\begin{equation}
\frac{d}{dq}
\left(\frac{q}{1+q+q^2}\right)
=\frac{1-q^2}{(1+q+q^2)^2},
\label{eq:concurrence-ratio-derivative}
\end{equation}
the concurrence~\eqref{eq:stationary-concurrence-general} is nonincreasing up to population balance and
nondecreasing beyond it.
We classify its acceleration dependence according to the conserved
singlet population $p_s$, at fixed coupling asymmetry $q_{\chi}$.

\subsubsection{Pure singlet: $p_s=1$}

The initial state is the pure singlet $\rho(0)=|s\rangle\langle s|$.
This is a maximally entangled dark state of the collective
coupling~\cite{Dicke54,Lehmberg70}: all collective coupling operators
annihilate it, $S_k|s\rangle=0$, and the effective
Hamiltonian~\eqref{eq:hamiltonian-singlet-triplet} preserves it.
Consequently, $\rho(\tau)=|s\rangle\langle s|$ at all times and
\begin{equation}
C[\rho(\tau)]=C[\hat\rho]=1.
\label{eq:pure-singlet-concurrence}
\end{equation}
This protection holds at any acceleration and for every coupling
compatible with the assumed collective structure.

 \subsubsection{No singlet population: $p_s=0$}

Here the concurrence vanishes for every acceleration and every
allowed coupling. All stationary states in this family are separable:
\begin{equation}
C[\hat\rho]=0.
\label{eq:no-singlet-concurrence}
\end{equation}

\subsubsection{Entanglement at every acceleration: $1/2<p_s<1$}

\begin{figure}[h]
\centering
\includegraphics[width=\linewidth]{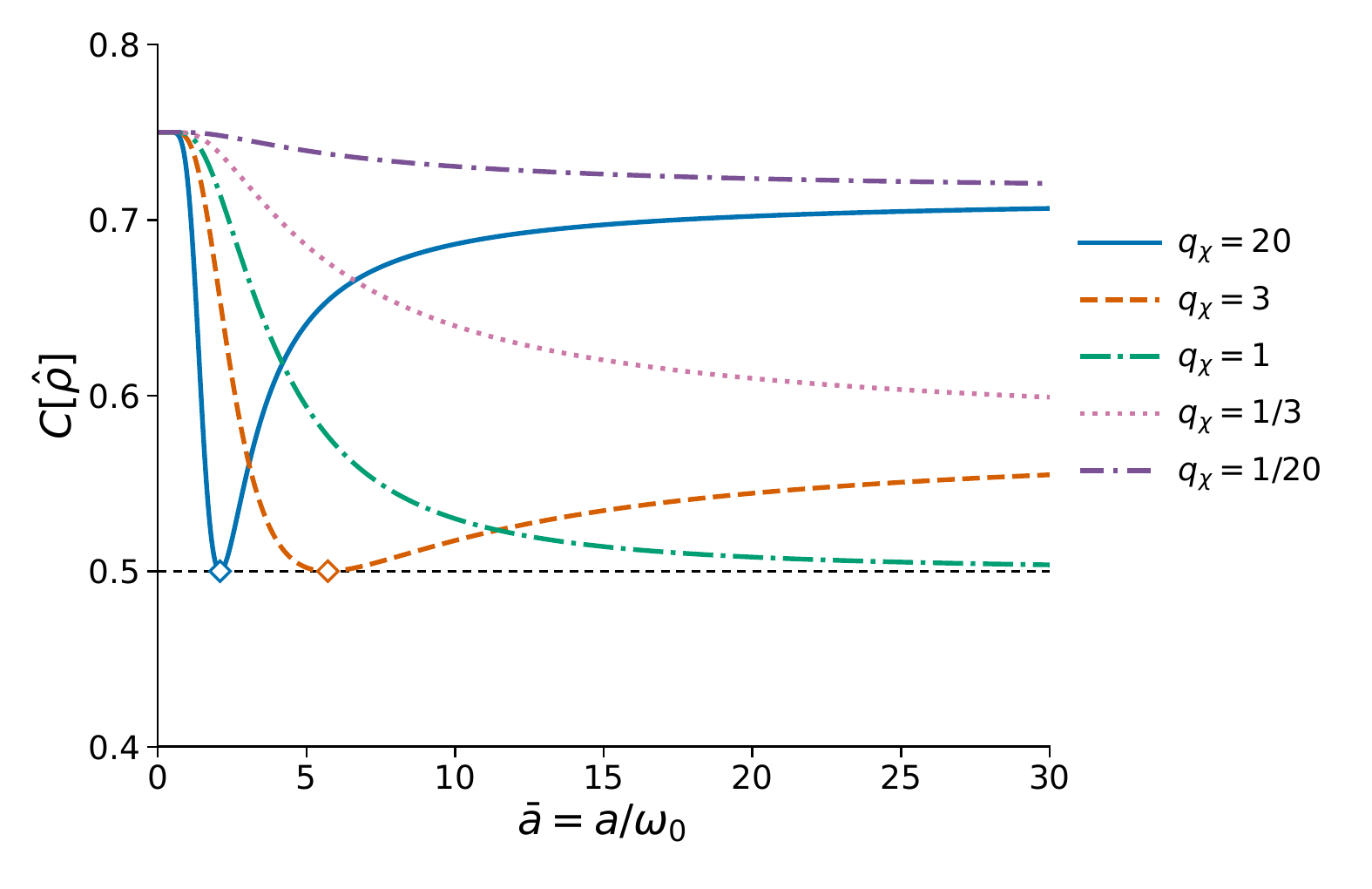}
\caption{Stationary concurrence for singlet population $p_s=3/4$, as a function of acceleration
$\acc=a/\om$, for coupling asymmetries $q_{\chi}=1/20,1/3,1,3,20$.
For $q_{\chi}\leq1$, the concurrence decreases monotonically.
For $q_{\chi}>1$, it reaches its minimum at population balance,
at $\acc_{\mathrm{inv}}$ in Eq.~\eqref{eq:inversion}, marked by diamonds,
and then increases after inversion. The dashed horizontal line marks
$C[\hat\rho]=2p_s-1=1/2$: the stationary state remains entangled
throughout. Each curve approaches the asymptotic value in
Eq.~\eqref{eq:positive-large-acceleration-limit}.}
\label{fig:classification-persistent}
\end{figure}

Equation~\eqref{eq:stationary-concurrence-general} gives
\begin{equation}
  C[\hat\rho]\geq2p_s-1>0,  
\end{equation}
so the stationary state remains entangled
at every acceleration.
For $0<q_{\chi}\leq1$, by~\eqref{eq:thermal-stationary-ratio} the concurrence decreases monotonically
towards
\begin{equation}
\lim_{\acc\to+\infty}C[\hat\rho]
=p_s-\frac{3(1-p_s)q_{\chi}}
{1+q_{\chi}+q_{\chi}^2}>0.
\label{eq:positive-large-acceleration-limit}
\end{equation}
For $q_{\chi}=1$, the lower bound $2p_s-1$ is approached only
as $\acc\to+\infty$; for $0<q_{\chi}<1$, the asymptotic value lies
above it. See Fig.~\ref{fig:classification-persistent}.

For $q_{\chi}>1$, the concurrence first decreases and then
increases. Its minimum occurs at population balance,
$q=1$, at
\begin{equation}
\acc_{\mathrm{inv}}=\frac{2\pi}{\ln(q_{\chi})}.
\label{eq:inversion}
\end{equation}
Each triplet population is then $(1-p_s)/3$, giving
\begin{equation}
C[\hat\rho]\big|_{\acc=\acc_{\mathrm{inv}}}=2p_s-1>0.
\label{eq:concurrence-at-inversion}
\end{equation}
Above this acceleration, $q>1$ and the concurrence rises
towards the limit~\eqref{eq:positive-large-acceleration-limit}.
Population inversion therefore occurs without loss of entanglement.
See Fig.~\ref{fig:classification-persistent}.

\subsubsection{Separability only at population balance: $p_s=1/2$}

\begin{figure}[h]
\centering
\includegraphics[width=\linewidth]{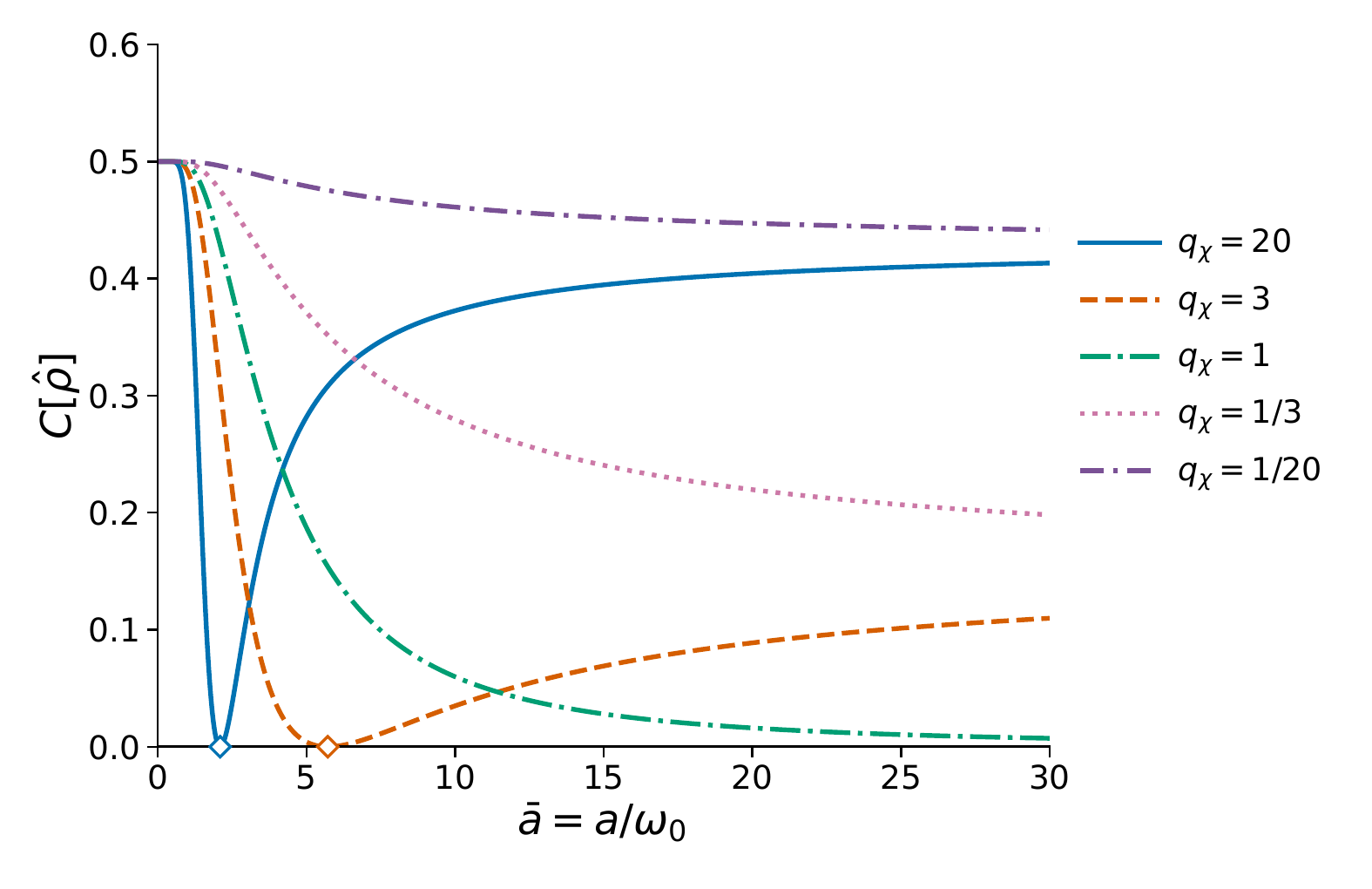}
\caption{Stationary concurrence for singlet population $p_s=1/2$, as a function
of acceleration $\acc=a/\om$, for coupling asymmetries $q_{\chi}=1/20,1/3,1,3,20$.
Each value $q_{\chi}>1$ yields a single separable point at
$\acc_{\mathrm{inv}}=2\pi/\ln(q_{\chi})$, marked by a diamond,
with positive concurrence on either side. For $q_{\chi}=1$,
the concurrence tends to zero only as $\acc\to+\infty$; for
$0<q_{\chi}<1$, it decreases towards the positive limit~\eqref{eq:half-singlet-large-acceleration-limit}.}
\label{fig:classification-balance}
\end{figure}

The concurrence~\eqref{eq:stationary-concurrence-general} reduces to
\begin{equation}
C[\hat\rho]=\frac{(1-q)^2}{2(1+q+q^2)}.
\label{eq:half-singlet-concurrence}
\end{equation}
It vanishes only at $q=1$, when the excitation and decay
rates and the three triplet populations are equal.
For $q_{\chi}>1$, this occurs at $\acc_{\mathrm{inv}}$.
The stationary state is separable only at that acceleration and
entangled on either side, without a finite separable interval.
For $q_{\chi}=1$, the concurrence decreases monotonically
towards zero, which is reached only as $\acc\to+\infty$.
For $0<q_{\chi}<1$, it also decreases monotonically but
approaches the strictly positive limit
\begin{equation}
\lim_{\acc\to+\infty}C[\hat\rho]
=\frac{(1-q_{\chi})^2}
{2(1+q_{\chi}+q_{\chi}^2)}>0.
\label{eq:half-singlet-large-acceleration-limit}
\end{equation}
See Fig.~\ref{fig:classification-balance}.

\subsubsection{Possible loss and recovery: $0<p_s<1/2$}
The boundaries of the separable interval are determined
by the vanishing of the expression inside the maximum
in the expression of the concurrence~\eqref{eq:stationary-concurrence-general}:
\begin{equation}
p_sq^2+(4p_s-3)q+p_s=0.
\end{equation}
Its two roots are reciprocal. Denote the smaller root by
\begin{equation}
q_{*}=
\frac{2p_s}{3-4p_s+\sqrt{3(1-2p_s)(3-2p_s)}},
\qquad 0<q_{*}<1.
\label{eq:critical-rate-ratio}
\end{equation}
The concurrence is positive precisely when
$q<q_{*}$ or $q>q_{*}^{-1}$.
The stationary state is therefore separable for
$q_{*}\leq q\leq q_{*}^{-1}$.

As acceleration increases, $q$ grows from zero towards
$q_{\chi}$. Entanglement is first lost when
$q=q_{*}$, at
\begin{equation}
\acc_{\mathrm{loss}}=
\frac{2\pi}{\ln(q_{\chi}/q_{*})}.
\label{eq:loss}
\end{equation}
This acceleration is finite and positive if and only if
$q_{\chi}>q_{*}$.
The position of $q_{\chi}$ relative to
$q_{*}$ and $q_{*}^{-1}$ distinguishes three regimes.

\begin{figure}[h]
\centering
\includegraphics[width=\linewidth]{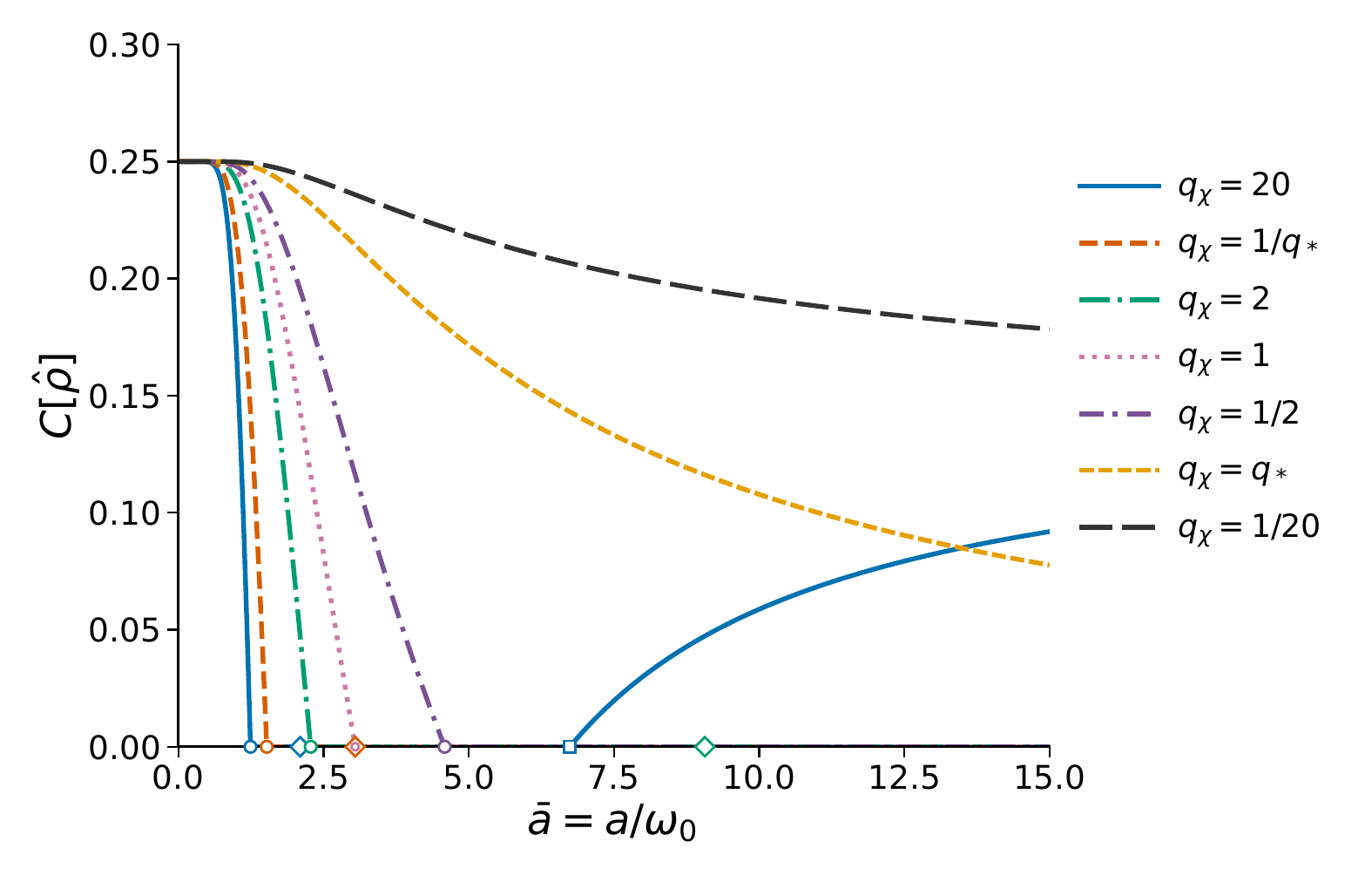}
\caption{Stationary concurrence for $p_s=1/4$, as a function
of acceleration $\acc=a/\om$.
The curves at $q_{\chi}=1/20$ and $q_{\chi}=q_{*}$, with $q_{*}=4-\sqrt{15}\simeq 0.127$,
remain entangled at every finite acceleration, although the latter
tends to zero as $\acc\to+\infty$.
For $q_{\chi}=1/2,1$, entanglement is lost without population
inversion; for $q_{\chi}=2,1/q_{*}$, loss is
followed by inversion without recovery.
For $q_{\chi}=20$, entanglement is recovered beyond a finite
separable interval. Circles, diamonds and squares mark loss,
population balance and recovery, respectively.}
\label{fig:classification-recovery}
\end{figure}

If $0<q_{\chi}\leq q_{*}$, the concurrence decreases
but remains positive at every finite acceleration.
For $q_{\chi}=q_{*}$, it approaches zero only as
$\acc\to+\infty$; for smaller values its limiting value is positive.
See Fig.~\ref{fig:classification-recovery}.

If the coupling asymmetry is in the interval $q_{*}<q_{\chi}\leq q_{*}^{-1}$, the concurrence
decreases to zero at $\acc_{\mathrm{loss}}$ in~\eqref{eq:loss} and remains zero thereafter.
For $q_{\chi}\leq1$, population inversion does not occur.
For $q_{\chi}>1$, inversion occurs above $\acc_{\mathrm{inv}}$ in~\eqref{eq:inversion},
but entanglement does not return because $q$ never exceeds
$q_{*}^{-1}$. At $q_{\chi}=q_{*}^{-1}$,
that second threshold is approached only as $\acc\to+\infty$.

Finally, for coupling asymmetries $q_{\chi}>q_{*}^{-1}$, the concurrence
decreases to zero, remains zero over a finite interval, and then
increases again. The upper boundary is reached at
$q=q_{*}^{-1}$, giving
\begin{equation}
\acc_{\mathrm{rev}}=
\frac{2\pi}{\ln(q_{\chi}q_{*})}.
\label{eq:recovery}
\end{equation}
The stationary state is separable for accelerations in the interval
$\acc_{\mathrm{loss}}\leq \acc\leq \acc_{\mathrm{rev}}$ and entangled again
for $\acc>\acc_{\mathrm{rev}}$.
Since the transition-rate ratio $q$ crosses $q_{*}$, $1$ and
$q_{*}^{-1}$ in this order, one has
$\acc_{\mathrm{loss}}<\acc_{\mathrm{inv}}<\acc_{\mathrm{rev}}$.
Population inversion occurs within the separable interval;
entanglement returns only when $q>q_{*}^{-1}$.
See Fig.~\ref{fig:classification-recovery}.

\begin{figure}[ht]
\centering
\includegraphics[width=0.85\columnwidth]{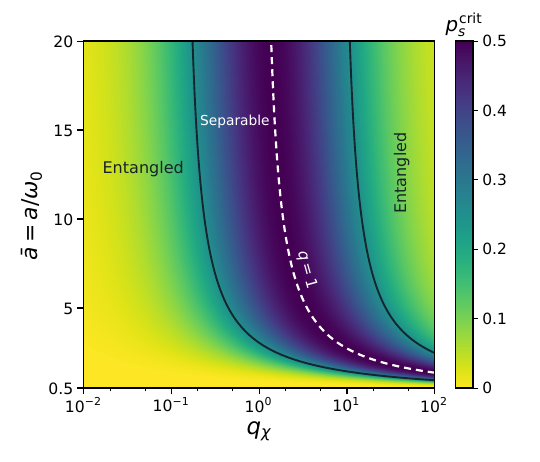}
\caption{Stationary-entanglement phase diagram for singlet population $p_s=1/4$ in the
coupling asymmetry-acceleration plane $(q_{\chi},\acc)$, with $\acc=a/\om$ and a logarithmic horizontal
axis. The color scale represents $p_s^{\mathrm{crit}}(q_{\chi},\acc)$.
The stationary state is entangled where $p_s^{\mathrm{crit}}<1/4$
and separable otherwise. Solid curves mark the boundaries
$q_{\chi}=q_{*}\mathrm{e}^{2\pi/\acc}$ and $q_{\chi}=q_{*}^{-1}\mathrm{e}^{2\pi/\acc}$, with
$q_{*}=4-\sqrt{15}$, where concurrence
vanishes. The white dashed curve marks population balance,
$q_{\chi}=\mathrm{e}^{2\pi/\acc}$, where the transition-rate ratio $q=1$ and
$p_s^{\mathrm{crit}}=1/2$. The endpoints of zero excitation or zero
decay coupling, corresponding to $q_{\chi}=0$ and
$q_{\chi}\to+\infty$, lie outside the plotted range and both
give a concurrence $C[\hat\rho]=p_s$.}
\label{fig:phase}
\end{figure}

Finally, we construct a phase diagram 
as the
coupling asymmetry $q_{\chi}$ and acceleration $\acc=a/\om$ vary.
Solving the condition of vanishing concurrence for $p_s$ in~\eqref{eq:stationary-concurrence-general} defines
the critical singlet population
\begin{equation}
p_s^{\mathrm{crit}}(q_{\chi},\acc)
=\frac{3q}{1+4q+q^2},
\qquad
q=q_{\chi}\,\mathrm{e}^{-2\pi/\acc}.
\label{eq:phase-threshold}
\end{equation}
Thus the stationary state $\hat{\rho}$ in~\eqref{eq:stationary-triplet} is entangled precisely when
\begin{equation}
p_s>p_s^{\mathrm{crit}}(q_{\chi},\acc),
\label{eq:phase-singlet-threshold}
\end{equation}
and separable otherwise.
Since $0\leq p_s^{\mathrm{crit}}\leq1/2$, every preparation with
$p_s>1/2$ yields an entangled stationary state throughout the domain.
See Fig.~\ref{fig:phase} for the case $p_s=1/4$. 

\subsection{Isotropic coupling tensor}
\label{sec:symmetric-coupling}
We now consider independent massless scalar fields
in four spacetime dimensions and choose isotropic
couplings. This is the situation that was considered in Ref.~\cite{Benatti04},
\begin{equation}
\chi_{rs}=\delta_{rs}.
\label{eq:isotropic-coupling-tensor}
\end{equation}
The detectors have the same orientation,
$\bm n^{(1)}=\bm n^{(2)}=\bm n$.
By isotropy, the results are independent of the common axis.
Since $\operatorname{Tr}Q=2$ and
$\operatorname{Tr}\Omega=0$,
Eq.~\eqref{eq:asymptotic-rate-ratio} gives for the coupling asymmetry $q_{\chi}=1$.

The field response function of a massless scalar field in $n=4$ spacetime dimensions is obtained in Appendix~\ref{app:response-functions} and reads
\begin{equation}
F_4(\omega)
=\frac{\omega}{2\pi\left(\mathrm{e}^{\beta\omega}-1\right)}
\,.
\label{eq:massless-response-4}
\end{equation}

Therefore, Eqs.~\eqref{eq:three-rates-tensors}
become
\begin{equation}
\begin{split}
\gamma_\downarrow
&=2F_4(-\om)
=\frac{\om}{\pi(1-\mathrm{e}^{-\beta\om})},\\
\gamma_\uparrow
&=2F_4(\om)
=\frac{\om}{\pi(\mathrm{e}^{\beta\om}-1)},\\
\gamma_0
&=F_4(0)
=\frac{1}{2\pi\beta}
=\frac{a}{4\pi^2}.
\end{split}
\label{eq:symmetric-rates}
\end{equation}
The eigenvalues
of the Kossakowski matrix~\eqref{eq:equal-orientation-gamma} are
$\gamma_\downarrow/2$, $\gamma_\uparrow/2$ and $\gamma_0$.
They are strictly positive at every finite $a>0$.
Hence, $\operatorname{rank}\Gamma=3$, and the assumed
convergence condition~\eqref{eq:rank-condition} is satisfied.

The transition-rate ratio is
\begin{equation}
q
=\frac{\gamma_\uparrow}{\gamma_\downarrow}
=\mathrm{e}^{-\beta\om}.
\label{eq:symmetric-rate-ratio}
\end{equation}
As acceleration increases, $q$ grows from zero towards one,
and the triplet populations approach balance.
The stationary concurrence follows directly from
Eq.~\eqref{eq:stationary-concurrence-general}, with the thermal ratio
in Eq.~\eqref{eq:symmetric-rate-ratio}. Acceleration leaves the singlet
population unchanged but increases the central-triplet population $p_0$.

\begin{figure}[h]
\centering
\includegraphics[width=\linewidth]{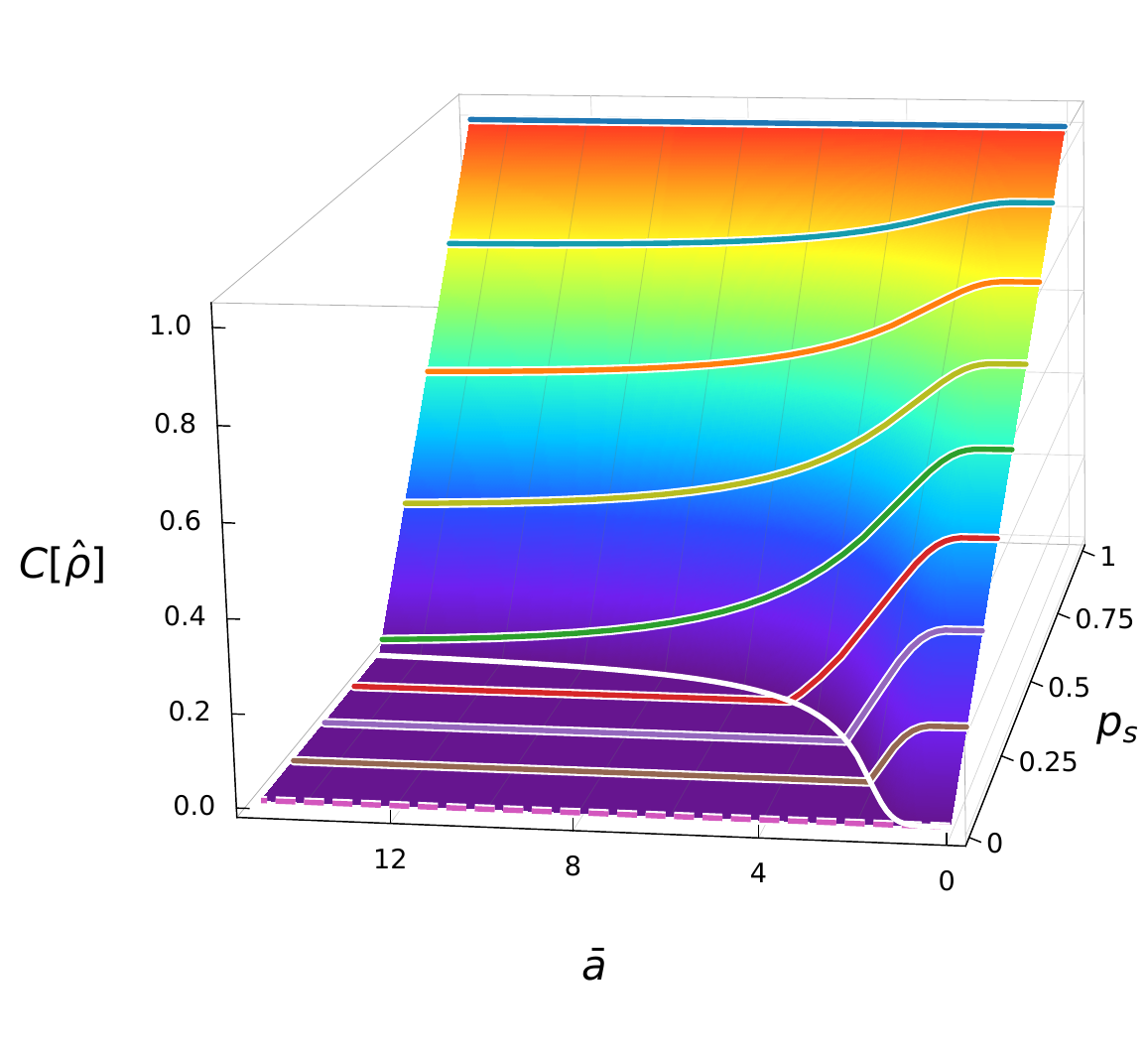}
\caption{Stationary concurrence for isotropic coupling
($q_{\chi}=1$), as a function of the acceleration $\acc=a/\om$
and the singlet population $p_s$.
From back to front, the solid colored curves correspond to
$p_s=1,\,7/8,\,3/4,\,5/8,\,1/2,\,3/8,\,1/4,\,1/8$,
and the dashed magenta line corresponds to $p_s=0$.
The white curve 
marks the boundary between positive and zero concurrence; see~\eqref{eq:phase-threshold}.
For $0<p_s<1/2$, entanglement is lost at the finite critical
acceleration $a_{\mathrm{loss}}$ in~\eqref{eq:symmetric-critical-acceleration}
and remains absent above it. For $1/2<p_s<1$, the concurrence
approaches the positive value $2p_s-1$, while for $p_s=1/2$
it vanishes only in the infinite-acceleration limit.}
\label{fig:concurrence3D}
\end{figure}

Figure~\ref{fig:concurrence3D} summarizes the dependence
of stationary concurrence on the dimensionless acceleration
$\acc$ and the conserved singlet population $p_s$.
At fixed $p_s$, the concurrence is nonincreasing with
acceleration, while increasing $p_s$ at fixed acceleration
enhances stationary entanglement once the separability
threshold is crossed.

In particular, for $0<p_s<1/2$, entanglement is present only for
$q<q_{*}$, with $q_{*}$ given in~\eqref{eq:critical-rate-ratio}. 
Entanglement is  lost when
$q=q_{*}$, at the acceleration value~\eqref{eq:loss},
\begin{equation}
a_{\mathrm{loss}}=\frac{2\pi\om}{\ln(1/q_{*})}.
\label{eq:symmetric-critical-acceleration}
\end{equation}
The concurrence decreases to zero at $a=a_{\mathrm{loss}}$, and the stationary
state remains separable for all $a\geq a_{\mathrm{loss}}$.
Isotropic coupling cannot produce population inversion or entanglement
recovery because $q<1$ at every finite acceleration.

For $1/2<p_s<1$, the concurrence instead decreases towards
\begin{equation}
\lim_{a\to+\infty}C[\hat\rho]=2p_s-1>0.
\label{eq:symmetric-large-acceleration-limit}
\end{equation}
At population balance $p_s=1/2$, it is given by~\eqref{eq:half-singlet-concurrence}
so it remains positive at every finite acceleration and vanishes only
as $a\to+\infty$. The pure singlet, $p_s=1$, retains unit concurrence,
whereas $p_s=0$ gives a separable stationary state throughout.

At low acceleration, the transition-rate ratio~\eqref{eq:symmetric-rate-ratio} is
exponentially small, and the triplet population concentrates in
$|t_-\rangle$. Taking the stationary limit first gives
\begin{equation}
\lim_{a\to0}C[\hat\rho]=p_s.
\label{eq:symmetric-low-acceleration-limit}
\end{equation}
For each fixed $0<p_s\leq1$, expanding in $q$ yields
\begin{equation}
C[\hat\rho]
=p_s-3(1-p_s)\mathrm{e}^{-2\pi\om/a}
+O\bigl(\mathrm{e}^{-4\pi\om/a}\bigr).
\label{eq:lowacc_expansion}
\end{equation}
For $0<p_s<1$, the leading correction is negative and exponentially
small in $\om/a$, producing a nearly flat low-acceleration plateau.
For the pure singlet the correction vanishes and concurrence remains one.
The nearly flat dependence on acceleration visible at
small $\acc$ in Fig.~\ref{fig:concurrence3D} reflects
these exponentially small corrections.

These results concern stationary entanglement and do not require
an entangled initial preparation. For example, the separable state
$|eg\rangle$ has $p_s=1/2$ and evolves towards a stationary state
with positive concurrence~\eqref{eq:half-singlet-concurrence}
at every finite $a>0$.

\subsection{Coupling tensor with symmetric and antisymmetric components}
\label{sec:mixed-coupling}
We retain massless scalar fields in four spacetime
dimensions and equal detector orientations, but
now choose
\begin{equation}
\chi_{jk}=\delta_{jk}-i \varepsilon_{jkl} n_l.
\label{eq:mixed-coupling-tensor}
\end{equation}
Thus,
$\chi=(I_3-i\Omega)$.
Using $\Omega^2=-Q$, the transverse coupling ratio~\eqref{eq:asymptotic-rate-ratio} is
\begin{equation}
q_{\chi}=
\frac{\operatorname{Tr}[(Q+i\Omega)(I_3-i\Omega)]}
{\operatorname{Tr}[(Q-i\Omega)(I_3-i\Omega)]}=0.
\label{eq:mixed-contractions}
\end{equation}

The transition rates~\eqref{eq:three-rates-tensors} now read
\begin{equation}
\begin{split}
\gamma_\downarrow
&=4F_4(-\om)
=\frac{2\om}{\pi(1-\mathrm{e}^{-\beta\om})},\\
\gamma_\uparrow&=0,\\
\gamma_0&=F_4(0)=\frac{a}{4\pi^2}.
\end{split}
\label{eq:mixed-rates}
\end{equation}
The Kossakowski matrix in
Eq.~\eqref{eq:equal-orientation-gamma} therefore reduces to
\begin{equation}
\Gamma
=\frac{\gamma_\downarrow}{4}(Q-i\Omega)+\gamma_0P.
\label{eq:mixed-kossakowski-matrix}
\end{equation}
Its eigenvalues are $\gamma_\downarrow/2$, $0$ and
$\gamma_0$. Since $\gamma_\downarrow>0$ and $\gamma_0>0$
at every finite $a>0$, one has
$\operatorname{rank}\Gamma=2$, and the 
convergence condition~\eqref{eq:rank-condition} is satisfied.
The coupling therefore suppresses collective excitation even though the accelerated field response has nonzero spectral weight at both transition frequencies. Consequently, $q=q_{\chi}=0$ at every  acceleration. The remaining decay channel transfers all triplet population to $|t_-\rangle$, while the singlet population remains conserved. Setting $q=0$ in Eq.~\eqref{eq:stationary-triplet} gives
\begin{equation}
\hat\rho
=p_s|s\rangle\langle s|+ (1-p_s)|t_-\rangle\langle t_-|,
\label{eq:mixed-stationary-state}
\end{equation}
whose concurrence is
\begin{equation}
C[\hat\rho]=p_s .
\label{eq:mixed-concurrence}
\end{equation}
Thus, for each fixed $p_s$, 
the stationary concurrence is independent of acceleration. The pure singlet, $p_s=1$, 
retains unit concurrence; every preparation with $0<p_s<1$ 
leads to an entangled stationary state; and $p_s=0$ 
gives a separable stationary state.

The different  behavior found above can be understood by examining the microscopic interactions associated with the two coupling tensors $\chi$ in~\eqref{eq:total-coupling-tensor}. 

Choose the common detector axis as
$\bm n=(0,0,1)$, so that $S_3$ is longitudinal and
$S_\pm=(S_1\pm iS_2)/2$ raise and lower the detector energy.
We recall that
$\phi_c^{(+)}$ and $\phi_c^{(-)}$ denote the Minkowski
annihilation and creation parts of the fields, respectively.

It is convenient to collect the field-detector coupling coefficients $\chi_{k}^c$
in the field operators~\eqref{eq:field_operator}
into three $N$-dimensional coupling vectors 
\begin{equation}
  |\chi_k\rangle =(\chi_k^1,\ldots,\chi_k^N)^T \in\mathbb C^N.  
\end{equation}
Here $k=1,2,3$ labels the detector coupling channel, while the upper index $c=1,\ldots,N$ labels the independent fields $\phi_c$.
The total coupling tensor~\eqref{eq:total-coupling-tensor} is thus given by the Gram matrix
\begin{equation}
    \chi_{jk} = \langle\chi_j|\chi_k\rangle.
\end{equation}

First consider the isotropic coupling~\eqref{eq:isotropic-coupling-tensor}.
It can be realized using three independent fields and the 3-dimensional vectors
\begin{equation}
 |\chi_1\rangle=
 \begin{pmatrix}
     1\\ 0\\ 0
 \end{pmatrix},
 \quad
 |\chi_2\rangle=
 \begin{pmatrix}
     0\\ 1\\ 0
 \end{pmatrix},
 \quad
 |\chi_3\rangle=
 \begin{pmatrix}
     0\\ 0\\ 1
 \end{pmatrix}.   
\end{equation}
Substituting these coefficients into the interaction Hamiltonian~\eqref{eq:H_int} gives
\begin{equation}
H'=S_3\otimes\phi_3
+S_+\otimes(\phi_1-i\phi_2)+S_-\otimes(\phi_1+i\phi_2).
\label{eq:example-interaction-local-isotropic}
\end{equation}
This interaction is local in the  real scalar fields $\phi_k$  ($k=1,2,3$). 
The longitudinal term coupled to $\phi_3$ produces dephasing without directly changing the detector populations.
The transverse interaction contains rotating-wave terms
$S_+\otimes\phi_c^{(+)}$ and $S_-\otimes\phi_c^{(-)}$,
which pair detector excitation with field annihilation
and detector de-excitation with field creation.
It also contains counter-rotating terms
$S_+\otimes\phi_c^{(-)}$ and $S_-\otimes\phi_c^{(+)}$,
which create or annihilate detector and field excitations
together, as in a Rabi-type coupling~\cite{Rabi37}.
Here $c=1,2$.
 The collective dynamics can generate entanglement from suitable separable initial states, and its stationary concurrence generally depends on acceleration, as discussed in Sec.~\ref{sec:symmetric-coupling}.

Now consider the coupling tensor with symmetric and antisymmetric components~\eqref{eq:mixed-coupling-tensor}. A realization uses two independent fields and the coupling vectors 
\begin{equation}
 |\chi_1\rangle=
 \begin{pmatrix}
     1\\ 0
 \end{pmatrix},
 \quad
 |\chi_2\rangle=
 \begin{pmatrix}
     -i\\ 0
 \end{pmatrix},
 \quad
 |\chi_3\rangle=
 \begin{pmatrix}
     0\\ 1
 \end{pmatrix}.   
\end{equation}
The interaction Hamiltonian~\eqref{eq:H_int}  becomes
\begin{equation}
H'=S_3\otimes\phi_2
+2S_+\otimes\phi_1^{(+)}
+2S_-\otimes\phi_1^{(-)}.
\label{eq:example-interaction-nonlocal-rotating}
\end{equation}

The longitudinal term coupled to $S_3$ again produces dephasing and involves a field independent of the transverse one. The transverse interaction contains only rotating-wave terms, as in a Jaynes--Cummings-type coupling~\cite{JaynesCummings63}.

The Unruh effect reflects the response of an accelerated detector to quantum vacuum fluctuations. In the Minkowski mode description, excitation from the vacuum involves counter-rotating terms, which simultaneously excite the detector and create a field quantum. Acceleration allows the external agent maintaining the trajectory to supply the required energy, so these otherwise virtual processes can produce real excitations~\cite{Funai19}. In the second example, the transverse interaction contains only rotating-wave terms. Detector excitation is therefore accompanied by field annihilation, which cannot act on the vacuum, while the longitudinal coupling does not directly change detector populations. The vacuum-excitation channel is consequently absent. 

Rotating-wave models are ubiquitous in quantum optics and cavity quantum electrodynamics~\cite{JaynesCummings63,JayneCummingsbook2021} and are routinely used as a very effective description of a variety of physical situations~\cite{Burgarth2024tamingrotatingwave, Burgarth2022oneboundtorulethem, burgarth2026rotatingwavesecularapproximationsopen}.
When applied to local relativistic detector-field
interactions, however, the rotating-wave approximation
can introduce nonlocality and permit faster-than-light
signalling~\cite{Funai19}.
From the perspective of the Unruh effect, in our model it produces no acceleration-dependent signature on  stationary entanglement, although acceleration still affects the relaxation rates; see Eq.~\eqref{eq:mixed-rates}.

\section{Unequal orientations and transient entanglement}
\label{sec:unequal-orientations}
We now investigate how a difference between the detector orientations affects the entanglement supported by collective symmetry.
We consider the isotropic coupling~\eqref{eq:isotropic-coupling-tensor}
and independent massless scalar fields in four
spacetime dimensions. The detectors have the same energy gap and follow the same uniformly accelerated worldline, but their internal axes are
\begin{equation}
\begin{aligned}
\bm n^{(1)}&=(0,0,1),
\\
\bm n^{(2)}&=(\sin\theta,0,\cos\theta).
\end{aligned}
\label{eq:unequal-orientation-axes}
\end{equation}
We consider $0\leq\theta\leq\pi$ and finite accelerations $a>0$.
The misalignment enters the dynamics only through
\begin{equation}
\eta=\sin^2\Bigl(\frac{\theta}{2}\Bigr),
\label{eq:misalignment-parameter}
\end{equation}
with $\cos^2(\theta/2)=1-\eta$ and $\cos(\theta)=1-2\eta$.

\subsection{Master equation in the rotated frame}
\label{sec:unequal-rotated-frame}

The second axis is obtained from the first one through the rotation
\begin{equation}
\bm n^{(2)}=R_2 \bm n^{(1)}, \qquad R_2(\theta)=
\begin{pmatrix}
\cos(\theta) & 0 & \sin(\theta)\\
0 & 1 & 0\\
-\sin(\theta) & 0 & \cos(\theta)
\end{pmatrix},
\label{eq:local-axis-rotations}
\end{equation}
which acts on the second detector through the local unitary
\begin{equation}
U_\theta=\sigma_0\otimes\mathrm{e}^{-i\theta\sigma_2/2},
\qquad
 U_\theta^\dagger\,\sigma_k^{(2)}\, U_\theta
=\sum_{l=1}^{3} (R_2)_{kl}\,\sigma_l^{(2)}.
\label{eq:rotated-frame}
\end{equation}
In particular, 
\begin{equation}
    U_\theta^\dagger(\bm n^{(2)}\cdot\bm\sigma^{(2)})\, U_\theta=\sigma_3^{(2)}.
\end{equation}
In the rotated frame, 
\begin{equation}
    \tilde{\rho}= U_\theta^\dagger\rho\, U_\theta,
\end{equation}
both detectors are therefore quantized along $\bm{n}^{(1)}=(0,0,1)$ and the free Hamiltonian becomes $\tilde{H}_S=\frac{\om}{2}S_3$, as for equal orientations.
Since $U_\theta$ is local, 
\begin{equation}
    C[\tilde{\rho}]=C[\rho].
\end{equation}
We thus work with $\tilde{\rho}$ and with the singlet-triplet basis~\eqref{eq:singlet-triplet-basis}.

The rotation maps the Cartesian blocks~\eqref{eq:coherent-blocks} and~\eqref{eq:dissipative-blocks} into
\begin{equation}
    \tilde\Lambda_{ab}=R_a^T\Lambda_{ab}R_b, \qquad \tilde\Gamma_{ab}=R_a^T\Gamma_{ab}R_b,
\end{equation}
with $R_1=I_3$.
For $\chi=I_3$, the local blocks coincide with those for equal orientations, while the cross-detector blocks are reduced by the misalignment:
\begin{equation}
\begin{aligned}
\tilde\Gamma_{11}&=\tilde\Gamma_{22}=\Gamma,
&\quad
\tilde\Gamma_{12}&=\tilde\Gamma_{21}=(1-\eta)\,\Gamma-\eta\,\gamma_0P,
\\
\tilde\Lambda_{11}&=\tilde\Lambda_{22}=\Lambda,
&\quad
\tilde\Lambda_{12}&=\tilde\Lambda_{21}=(1-\eta)\,\Lambda-\eta\,J_\parallel P.
\end{aligned}
\label{eq:rotated-blocks}
\end{equation}
Here $\Gamma$ and $\Lambda$ are the equal-orientation matrices~\eqref{eq:equal-orientation-gamma} and~\eqref{eq:equal-orientation-lambda} for $\bm n=(0,0,1)$, with the rates~\eqref{eq:symmetric-rates}. In particular,
\begin{equation}
    \gamma_\uparrow=q\gamma_\downarrow, \qquad q=\mathrm{e}^{-\beta\om},
\end{equation}
with $\beta=2\pi / a$.
For $\eta=0$ all four blocks coincide,
$\tilde{\Gamma}_{ab}=\Gamma$ and $\tilde{\Lambda}_{ab}=\Lambda$
as for equal orientations.

Since $\tilde\Gamma_{11}=\tilde\Gamma_{22}$ and $\tilde\Gamma_{12}=\tilde\Gamma_{21}$, the dissipator~\eqref{eq:dissipative_term} separates into the collective operators $S_j$ and the antisymmetric combinations
\begin{equation}
\Sa_j=\sigma_j^{(1)}-\sigma_j^{(2)},
\qquad
\Sa_\pm=\frac12(\Sa_1\pm i\Sa_2),
\label{eq:antisymmetric-operators}
\end{equation}
with coefficient matrices $(\tilde\Gamma_{11}+\tilde\Gamma_{12})/2$ and $(\tilde\Gamma_{11}-\tilde\Gamma_{12})/2$, respectively.
Reducing each of them to three channels, as in Eq.~\eqref{eq:three-channels}, gives the master equation in the rotated frame,
\begin{equation}
\begin{split}
\dot{\tilde{\rho}}={}&-i\bigl[\tilde{H}_{\mathrm{eff}},\tilde{\rho}\bigr]
\\
&+\Bigl(1-\frac{\eta}{2}\Bigr)
\bigl(\gamma_\downarrow\mathcal{D}_{S_-}+\gamma_\uparrow\mathcal{D}_{S_+}\bigr)[\tilde{\rho}]
+(1-\eta)\,\gamma_0\mathcal{D}_{S_3}[\tilde{\rho}]\\
&+\frac{\eta}{2}
\bigl(\gamma_\downarrow\mathcal{D}_{\Sa_-}+\gamma_\uparrow\mathcal{D}_{\Sa_+}\bigr)[\tilde{\rho}]+\eta\,\gamma_0\mathcal{D}_{\Sa_3}[\tilde{\rho}],
\end{split}
\label{eq:misaligned-generator}
\end{equation}
with the rates~\eqref{eq:symmetric-rates}, and the effective Hamiltonian, up to an additive scalar,
\begin{equation}
\begin{split}
\tilde{H}_{\mathrm{eff}}={}&\frac{\om+\delta\omega}{2}S_3
+(1-2\eta)\,J_\parallel\,\sigma_3^{(1)}\sigma_3^{(2)}\\
&+(1-\eta)\,J_\perp\bigl(\sigma_1^{(1)}\sigma_1^{(2)}+\sigma_2^{(1)}\sigma_2^{(2)}\bigr).
\end{split}
\label{eq:misaligned-hamiltonian}
\end{equation}
The local part follows by plugging~\eqref{eq:rotated-blocks} into~\eqref{eq:H_a} and the cross-detector part from $\operatorname{Re}\tilde\Lambda_{12}$ through~\eqref{eq:H_12}. The coefficients 
\begin{equation}
   \delta\omega = - 2 K_-(\om), \quad 
   J_\parallel = K_+(0), \quad J_\perp= K_+(\om)
\end{equation}
are the dispersive coefficients~\eqref{eq:coupling_parameters} and~\eqref{eq:transverse-coherent-coupling} for $\chi=I_3$, where
\begin{equation}
\begin{aligned}
K_\pm(\Om)=\frac12\bigl[K_4(\Om)\pm K_4(-\Om)\bigr].
\end{aligned}
\label{eq:unequal-even-odd-functions}
\end{equation}
With the inertial-reference subtraction of
Appendix~\ref{app:lamb_shift}, the retained $K_4$ is odd,
so $K_+(\Om)=0$ and the cross-detector terms
in the Hamiltonian~\eqref{eq:misaligned-hamiltonian} vanish.

For $\eta=0$, Eqs.~\eqref{eq:misaligned-generator} and~\eqref{eq:misaligned-hamiltonian} reduce to the equal-orientation master equation, with the three collective channels~\eqref{eq:three-channels} and the Hamiltonian~\eqref{eq:collective-hamiltonian}.
Misalignment transfers a fraction of each channel, proportional to $\eta$, to the antisymmetric operators $\Sa_j$. Unlike the collective operators, these do not annihilate the singlet,
\begin{equation}
\Sa_\pm|s\rangle=\mp\sqrt2\,|t_\pm\rangle,
\qquad
\Sa_3|s\rangle=2\,|t_0\rangle.
\label{eq:antisymmetric-on-singlet}
\end{equation}
The singlet is therefore no longer a dark state, and its population is no longer conserved.

\subsection{Stationary state}
\label{sec:unequal-stationary-state}

In the rotated frame, $S_-$ and $\Sa_-$ lower the energy by $\om$, their adjoints raise it, and the corresponding rates in Eq.~\eqref{eq:misaligned-generator} are in the ratio $\gamma_\uparrow/\gamma_\downarrow=q=\mathrm{e}^{-\beta\om}$.
The Gibbs state of the free Hamiltonian therefore satisfies detailed balance channel by channel, while $S_3$ and $\Sa_3$ commute with it. This state factorizes,
\begin{equation}
\hat{\rho}_G
=\frac{\mathrm{e}^{-\beta\om S_3/2}}{\operatorname{Tr}\bigl(\mathrm{e}^{-\beta\om S_3/2}\bigr)}
=\rho_\beta\otimes\rho_\beta,
\label{eq:unequal-orientation-gibbs}
\end{equation}
with
\begin{equation}
    \rho_\beta=\frac{1}{1+q}\bigl(|g\rangle\langle g|+q\,|e\rangle\langle e|\bigr).
\end{equation}
In the original frame $\hat{\rho}_{\theta}= U_\theta\hat{\rho}_G\, U_\theta^\dagger$
is the Gibbs state 
of the free Hamiltonian $H_S$.

In the singlet--triplet basis, $\hat{\rho}_G$ is diagonal, with populations
\begin{equation}
\hat{p}_s=\frac{q}{(1+q)^2},
\qquad
(\hat{p}_-,\, \hat{p}_0, \, \hat{p}_+)=\frac{1}{(1+q)^2}(1,\, q,\, q^2).
\label{eq:gibbs-singlet-population}
\end{equation}
Since $1-\hat{p}_s=Z_{q}/(1+q)^2$, this is precisely the member of the equal-orientation family~\eqref{eq:stationary-triplet} with $p_s=\hat{p}_s$. It commutes with $\tilde{H}_{\mathrm{eff}}$, which is diagonal in the same basis, and is therefore stationary under the full generator. Moreover, $\hat{p}_s=\hat{p}_0<3\hat{p}_0$, so that its concurrence~\eqref{eq:stationary-concurrence-general} vanishes, consistently with its product form~\eqref{eq:unequal-orientation-gibbs}.

For $0<\theta\leq\pi$, i.e.\ $\eta>0$, all transition coefficients in Eq.~\eqref{eq:misaligned-generator} are strictly positive, and the jump operators generate the local raising and lowering operators 
\begin{equation}
 \sigma_\pm^{(1)}=\frac{1}{2}(S_\pm+\Sa_\pm), \qquad   \sigma_\pm^{(2)}=\frac{1}{2}(S_\pm-\Sa_\pm).
\end{equation}
An operator commuting with all jump operators and their adjoints must therefore commute with $\sigma_\pm^{(a)}$, $a=1,2$, and is proportional to $I_4$. Since $\hat{\rho}_G$ has full rank for $0<q<1$, Frigerio's criterion~\cite{Frigerio78} implies that it is the unique stationary state and attracts every initial preparation. Because it is a product state,
\begin{equation}
\lim_{\tau\to+\infty}C[\rho(\tau)] = C[\hat\rho_\theta] =0 \,,
\label{eq:unequal-orientation-zero-concurrence}
\end{equation}
for every $0<\theta\leq\pi$.

At $\theta=0$ the antisymmetric channels vanish, $p_s$ is conserved, and the stationary states form the family~\eqref{eq:stationary-triplet} studied in Sec.~\ref{sec:symmetric-coupling}, whose concurrence can be positive. Any misalignment selects from this family the single separable member~\eqref{eq:gibbs-singlet-population}.

\subsection{Transient entanglement and slow relaxation}
\label{sec:unequal-transient-entanglement}

To isolate entanglement generated by the dynamics, we use the same separable initial state 
\begin{equation}
    \rho(0)=\frac{I_4}{4}
    \label{eq:initial-state}
\end{equation}
 for every angle. It is invariant under $U_\theta$, so that $\tilde{\rho}(0)=I_4/4$.
The operators $\tilde{H}_{\mathrm{eff}}$, $S_3$ and $\Sa_3$ are diagonal in the singlet--triplet basis, and $S_\pm$ and $\Sa_\pm$ map basis vectors onto basis vectors. Hence $\tilde{\rho}(\tau)$ remains diagonal, 
\begin{align}
\tilde{\rho}(\tau)
={}& p_s(\tau)|s\rangle\langle s|
\nonumber\\
&+ p_-(\tau)|t_-\rangle\langle t_-|
+p_0(\tau)|t_0\rangle\langle t_0|+p_+(\tau)|t_+\rangle\langle t_+|,
\label{eq:diagonal-tilde-rho}
\end{align}
and its concurrence  has the form~\eqref{eq:singlet-triplet-concurrence} 
\begin{equation}
C[\tilde{\rho}(\tau)]=\max\Bigl\{0, \, |p_s(\tau)-p_0(\tau)|-2\sqrt{p_-(\tau)p_+(\tau)}\Bigr\},
\label{eq:singlet-triplet-concurrence-tau}
\end{equation}
with the time-dependent populations $p_s(\tau)$ and $p_m(\tau)$ ($m\in\{-,0,+\}$).
By plugging~\eqref{eq:diagonal-tilde-rho} into~\eqref{eq:misaligned-generator} and using~\eqref{eq:antisymmetric-on-singlet}, the triplet populations obey
\begin{equation}
\begin{split}
\dot p_-={}&2\gamma_\downarrow p_0-2\gamma_\uparrow p_-
+\eta\,\gamma_\downarrow(p_s-p_0),\\
\dot p_0={}&2\gamma_\downarrow(p_+-p_0)+2\gamma_\uparrow(p_--p_0)\\
&-\eta\bigl(\gamma_\downarrow(p_+-p_0)+\gamma_\uparrow(p_--p_0)
-4\gamma_0(p_s-p_0)\bigr),\\
\dot p_+={}&2\gamma_\uparrow p_0-2\gamma_\downarrow p_+
+\eta\,\gamma_\uparrow(p_s-p_0),
\end{split}
\label{eq:misalignment-triplet-populations}
\end{equation}
and the singlet population obeys
\begin{equation}
\begin{split}
\dot p_s
=\eta\bigl(\gamma_\uparrow p_-+\gamma_\downarrow p_++4\gamma_0p_0
-(\gamma_\uparrow+\gamma_\downarrow+4\gamma_0)\,p_s\bigr).
\end{split}
\label{eq:misalignment-singlet-population}
\end{equation}
For $\eta=0$, Eqs.~\eqref{eq:misalignment-triplet-populations} reduce to Eqs.~\eqref{eq:triplet-populations}, and $p_s$ is conserved.

For small angles, $\eta\simeq\theta^2/4$. Transitions within the triplet sector remain of order one, whereas all transitions involving the singlet are proportional to $\eta$. The triplet populations therefore relax first, at essentially fixed $p_s$, to the relative weights of Eq.~\eqref{eq:stationary-triplet},
\begin{equation}
\bigl(p_-(\tau),\, p_0(\tau),\, p_+(\tau)\bigr)
\simeq
\frac{1-p_s(\tau)}{Z_{q}}\,(1,\, q,\, q^2),
\label{eq:misalignment-conditional-triplets}
\end{equation}
where $Z_{q} = 1+q+q^2$.
Substituting into~\eqref{eq:misalignment-singlet-population}, and using $\gamma_\uparrow=q\gamma_\downarrow$, 
gives
\begin{equation}
\dot p_s(\tau)
\simeq
-r_\theta
\Bigl(p_s(\tau)-\frac{q}{(1+q)^2}\Bigr),
\label{eq:misalignment-slow-population}
\end{equation}
with
\begin{equation}
r_\theta=
\eta\,(\gamma_\downarrow+\gamma_\uparrow+4\gamma_0)
\frac{(1+q)^2}{1+q+q^2}
+O(\eta^2)
\label{eq:unequal-orientation-slow-rate}
\end{equation}
at fixed positive acceleration $a>0$.
The weak singlet transitions thus drive $p_s$ towards its Gibbs value~\eqref{eq:gibbs-singlet-population} on the long timescale $r_\theta^{-1}\propto\theta^{-2}$.

During this slow stage, the state remains close to the member of the equal-orientation family~\eqref{eq:stationary-triplet} with the instantaneous singlet population $p_s(\tau)$, which constitutes the metastable regime~\cite{MacieszczakEtAl2016}. Its concurrence is therefore approximately given by~\eqref{eq:stationary-concurrence-general}, that is
\begin{equation}
C[\tilde{\rho}(\tau)]
=\max\left\{0,p_s(\tau)-\frac{3q \bigl(1-p_s(\tau)\bigr)}{1+q+q^2}\right\}.
\label{eq:approx-concurrence-tau}
\end{equation}

Within the metastable approximation, entanglement
disappears when $p_s(\tau)$ falls below the critical value
\begin{equation}
p_s^{\mathrm{crit}}(1,\bar{a})
=\frac{3q}{1+4q+q^2}
\end{equation}
of~\eqref{eq:phase-threshold} with $q_{\chi}=1$.

For the initial condition~\eqref{eq:initial-state},
$p_s(0)=1/4$, a positive metastable concurrence $p_s(0)>p_s^{\mathrm{crit}}$ requires $q< q_*=4-\sqrt{15}$, that is,
by~\eqref{eq:symmetric-critical-acceleration},
\begin{equation}
0<\acc<\acc_{\mathrm{loss}}
\simeq 3.045.
\label{eq:transient-entanglement-condition}
\end{equation}
In this regime, for sufficiently small nonzero
$\theta$ and at fixed acceleration away from the
critical value, integrating
Eq.~\eqref{eq:misalignment-slow-population}
gives the leading-order estimate for the lifetime
of transient entanglement,
\begin{align}
\tau_\theta &\simeq\frac{1}{r_\theta}
\ln\!\left(\frac{p_s(0)-\hat{p}_{s}}{p_s^{\mathrm{crit}}-\hat{p}_{s}}\right)
\nonumber\\
&= \frac{1}{r_\theta} \ln \left(\frac{(1-q)^2(1+4q+q^2)}{8 q(1+q+q^2)}\right),
\label{eq:transient-lifetime}
\end{align}
which grows as $\theta^{-2}$ as alignment is approached.

\begin{figure}[h]
\centering
\includegraphics[width=\linewidth]{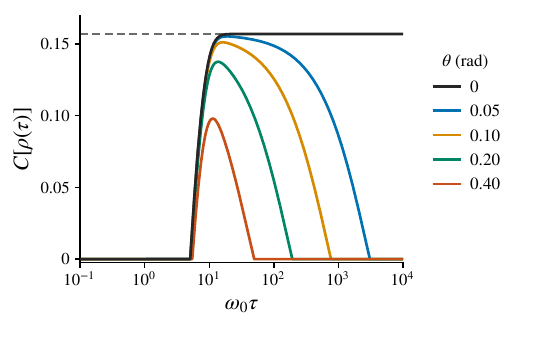}
\caption{Concurrence versus $\om\tau$ at $\bar{a}=a/\om=2$ for isotropic coupling to massless scalar fields in four spacetime dimensions, starting from $\rho(0)=I_4/4$. The solid black curve, corresponding to $\theta=0$, approaches the stationary value $C_*\simeq 0.157$ given by Eq.~\eqref{eq:concurrence-plateau}. The dashed horizontal line marks this asymptote. The colored solid curves, for $\theta=0.05,0.10,0.20,0.40$ rad, show transient entanglement that eventually disappears. The time axis is logarithmic.}
\label{fig:unequal-orientations}
\end{figure}

Figure~\ref{fig:unequal-orientations} shows this behaviour at $\acc=a/\om=2$.
At $\theta=0$, singlet conservation gives $p_s=1/4$,
while $\acc=2$ implies $q=\mathrm{e}^{-\pi}$.
Equation~\eqref{eq:stationary-concurrence-general}
therefore yields
\begin{equation}
C_*=\lim_{\tau\to \infty} C[\tilde{\rho}(\tau)]\bigl|_{\theta=0}
=\frac{\cosh(\pi)-4}{2\bigl(1+2\cosh(\pi)\bigr)}
\simeq0.157.
\label{eq:concurrence-plateau}
\end{equation}
For small nonzero angles, the concurrence first approaches this plateau and then vanishes at $\tau_\theta$; the smaller the angle, the closer and the longer the transient.
Here $p_s^{\mathrm{crit}}\simeq0.110$ and $\hat{p}_{s}\simeq0.040$, so that $\om \tau_{\theta} \simeq 7.62/\theta^2$.
At any fixed nonzero angle, waiting long enough gives the separable Gibbs state~\eqref{eq:unequal-orientation-gibbs}. Setting the angle to zero first restores singlet conservation and allows stationary entanglement.

\section{Discussion}
\label{sec:discussion}

We have derived a general master equation for two uniformly accelerated two-level atoms interacting with bosonic fields in the Minkowski vacuum.
Its generality lies in allowing arbitrary internal orientations, admissible coupling tensors, and any finite number of independent scalar fields with arbitrary masses, while considering both the coherent and dissipative contributions.
The spacetime dimension and field masses enter through the response functions and their Hilbert transforms.
The resulting generator therefore provides a common framework for investigating several physical scenarios: massive and massless environments in different spacetime dimensions, symmetric and asymmetric interactions, and aligned or misaligned internal axes.
These possibilities can be explored within the same formulation by specifying the field spectra and coupling coefficients.

For equal orientations, the condition $\operatorname{rank}(\Gamma)\geq2$
guarantees the exponential decay of all singlet-triplet coherences, so that
the stationary state is fixed by the conserved singlet population $p_s$; we
determined these states and quantified their entanglement through the
concurrence.

For fixed couplings and a common field spectrum, we classify the
acceleration dependence through the coupling asymmetry
$q_{\chi}$ and the conserved singlet population $p_s$.
Increasing acceleration can restore stationary entanglement after its
complete disappearance. For $0<p_s<1/2$, this occurs when
$q_{\chi}>q_{*}^{-1}$, with $q_{*}$ defined in
Eq.~\eqref{eq:critical-rate-ratio}: concurrence decreases to zero,
remains zero over a finite interval, and becomes positive again beyond
the recovery threshold. Recovery follows population inversion and
occurs when $q>q_{*}^{-1}$.
For $q_{\chi}>1$, a decrease followed by an increase also occurs
when $1/2<p_s<1$, although concurrence remains strictly positive;
at $p_s=1/2$, it vanishes only at population balance.
The recovery of stationary entanglement requires a sufficiently asymmetric transverse coupling. Within our scalar-field model, such asymmetry is realized by complex coupling coefficients, which generally correspond to nonlocal effective field observables.

For isotropic coupling, $q_{\chi}=1$, the transition-rate ratio
is fixed by the thermal response, and stationary concurrence is a
nonincreasing function of acceleration~\cite{Benatti04}.
For $0<p_s<1/2$, the critical acceleration is
$a_{\mathrm{loss}}=2\pi\om/\ln(1/q_{*})$, and the stationary state remains
separable for all $a\geq a_{\mathrm{loss}}$. This is the isotropic case of the
general classification: a sufficiently large asymptotic ratio,
$q_{\chi}>q_{*}^{-1}$, permits subsequent recovery.
For $1/2<p_s<1$, the isotropic model retains positive concurrence at
every acceleration, while for $p_s=1/2$ it approaches zero only at
infinite acceleration. The pure singlet remains maximally entangled,
whereas preparations with no singlet population yield stationary separable states.

The coupling with symmetric and antisymmetric components provides an explicit realization of acceleration-independent stationary entanglement.
For the coupling in Eq.~\eqref{eq:mixed-coupling-tensor},
the interaction contains only rotating-wave terms in the Minkowski mode decomposition: atomic excitation is accompanied by field annihilation, which cannot act on the vacuum.
The counter-rotating processes responsible for vacuum excitation are absent~\cite{Soda22}, while the longitudinal interaction produces dephasing without directly changing the populations.
Consequently, all triplet population relaxes to $|t_-\rangle$, and the stationary concurrence is $C[\hat\rho]=p_s$, 
independently of acceleration.
The stationary entanglement therefore carries no acceleration-dependent signature of Unruh excitation, even though the relaxation rates remain acceleration dependent.
This realization belongs to the effective nonlocal interactions obtained by selecting the Minkowski frequency components~\cite{Funai19}.

Finally, unequal internal orientations reveal the role of alignment in supporting stationary entanglement.
For the local isotropic model with massless fields in four spacetime dimensions, every nonzero misalignment at fixed positive acceleration removes singlet-population conservation and yields a unique separable product Gibbs state.
Nevertheless, small misalignment allows the dynamics to generate long-lived entanglement from the initially separable state $I_4/4$ in the regime examined.
We find that the slow population relaxation rate scales as $\theta^2$, so the associated relaxation time grows as $\theta^{-2}$ near alignment.
The resulting separation between rapid triplet relaxation and slow singlet relaxation produces an entangled metastable regime~\cite{MacieszczakEtAl2016}.
For these trajectories, the effective Hamiltonian commutes with the evolving density matrix, making their concurrence independent of the dispersive coefficients.
Exact alignment therefore supports stationary entanglement, while small deviations replace it with transient entanglement whose persistence is controlled by the misalignment angle.

\begin{acknowledgments}
The authors acknowledge the use of ChatGPT OpenAI for assistance with language editing, checking the consistency of the notation and exposition, and improving the clarity and organization of the manuscript. All AI-generated suggestions were reviewed and verified by the authors, who take full responsibility for the content.
We acknowledge support from the Italian National Group of Mathematical Physics (GNFM-INdAM), from INFN through the project ``QUANTUM", and from the Italian funding within the ``Budget MUR - Dipartimenti di Eccellenza 2023--2027" - Quantum Sensing and Modelling for One-Health (QuaSiModO).
PF acknowledges support from PNRR MUR project CN00000013-``Italian National Centre on HPC, Big Data and Quantum Computing``. 
PF acknowledges support from Projects PN-RIC Q-SUD (Q-SUD B99H26000380007) and PN-RIC QUANTAS (SYNERGIA B99H26000410007).

\end{acknowledgments}

\appendix

\section{Derivation and Cartesian form of the master equation}
\label{app:derivationGKLS}

The frequency components of the collective operators are
\begin{equation}
S_{k}^{(\Om)}=\sigma_{k}^{(\Om,1)}\otimes\sigma_{0}+\sigma_{0}\otimes\sigma_{k}^{(\Om,2)} \, .
\end{equation}
For each detector $a=1,2$, the operators $\sigma_r^{(\Om,a)}$ are defined through the projections $P^{\pm}_{\bm n^{(a)}}$ onto the eigenspaces of the corresponding single-detector Hamiltonian in Eq.~\eqref{eq:system_hamiltonian}:
\begin{equation}
P^{\pm}_{\bm n^{(a)}} 
= \frac{1}{2}\bigl(\sigma_0 \pm \bm n^{(a)}\cdot\bm\sigma\bigr) \, ,
\end{equation}
such that
\begin{eqnarray}
    \sigma_{k}^{(0,a)} &=& P^{+}_{\bm n^{(a)}} \sigma_{k} P^{+}_{\bm n^{(a)}}
+ P^{-}_{\bm n^{(a)}}\sigma_{k}P^{-}_{\bm n^{(a)}} \, , 
\label{eq:sigma_zeroapp}\\
\sigma_{k}^{(\pm \om,a)} &=& P^{\pm}_{\bm n^{(a)}} \sigma_{k} P^{\mp}_{\bm n^{(a)}} \,,
\label{eq:sigma_pmapp}
\end{eqnarray}
with $k = 1,2,3$.
In terms of the Pauli matrices, Eqs.~\eqref{eq:sigma_zeroapp} and~\eqref{eq:sigma_pmapp} become
\begin{equation}
\begin{split}
\sigma^{(0,a)}_{i}&= 
(P_a)_{ij}\, \sigma_{j} \,, \\
\sigma^{(\pm\om, a)}_{i}&=
\frac{1}{2} (Q_a \pm i \Omega_a)_{ij}\, \sigma_{j} \,,\qquad (i=1,2,3), 
\end{split}
\end{equation}
where the matrices are defined in Eq.~\eqref{eq:detector-geometric-matrices}.

Using Eqs.~\eqref{eq:responsefunc} and~\eqref{eq:Hilberttransf}, we obtain
\begin{equation}
\begin{split}
\alpha_{rs}(\Om)&=\int_{-\infty}^{\infty} \mathrm{d}\tau \, \mathrm{e}^{-i \Om \tau}\langle 0_M| \Phi_{r}\bigl(x(\tau)\bigr) \Phi_{s}\bigl(x(0)\bigr)|0_M\rangle  \\
&=\sum_{c=1}^{N}(\chi_{r}^{c})^*  \chi_{s}^{c}\, F_{n}^{(c)}(\Om) \,
\label{eq:alphaapp}
\end{split}
\end{equation}
and
\begin{equation}
\begin{split}
\kappa_{rs}(\Om)&=-i \int_{0}^{\infty} \mathrm{d} \tau\, \mathrm{e}^{-i \Om \tau}\langle 0_M| \Phi_{r}\bigl(x(\tau)\bigr) \Phi_{s}\bigl(x(0)\bigr)|0_M\rangle  \\
&+i\int_{0}^{\infty} \mathrm{d} \tau\, \mathrm{e}^{i \Om \tau}\langle 0_M| \Phi_{r}\bigl(x(0)\bigr) \Phi_{s}\bigl(x(\tau)\bigr)|0_M\rangle  \\
&=\sum_{c=1}^{N}\,(\chi_{r}^{c})^*  \chi_{s}^{c} \frac{\mathcal{P}}{\pi} \int_{-\infty}^{\infty} \mathrm{d} z \frac{F_{n}^{(c)}(z)}{z-\Om}  \\
&=\sum_{c=1}^{N} \, (\chi_{r}^{c})^*  \chi_{s}^{c} \,  K_{n}^{(c)}(\Om)\, .
\label{eq:betaapp}
\end{split}
\end{equation}

We derive the block expressions in
Eqs.~\eqref{eq:coherent-blocks} and~\eqref{eq:dissipative-blocks}
and record the expanded forms of Eqs.~\eqref{eq:H_a}, \eqref{eq:H_12}, and~\eqref{eq:dissipative_term}.
Using the geometric matrices defined in
Eq.~\eqref{eq:detector-geometric-matrices}, introduce 
\begin{equation}
\psi_a^{(0)}=P_a,
\,
\psi_a^{(\pm\om)}=\frac12(Q_a\pm i\Omega_a).
\label{eq:appendix-geometric-projectors}
\end{equation}
 These matrices are Hermitian. 
Equations~\eqref{eq:jumps2} 
therefore yield
\begin{equation}
\begin{split}
\Gamma_{ab}
&=\sum_{c=1}^{N}\sum_{\Om=0,\pm\om}
F_n^{(c)}(\Om)\psi_a^{(\Om)}\chi^{(c)}\psi_b^{(\Om)},\\
\Lambda_{ab}
&=\sum_{c=1}^{N}\sum_{\Om=0,\pm\om}
K_n^{(c)}(\Om)\psi_a^{(\Om)}\chi^{(c)}\psi_b^{(\Om)}.
\end{split}
\label{eq:appendix-spectral-blocks}
\end{equation}

For the Cartesian expressions, we can define the following coefficients determined by the response functions $F_n^{(c)}(\om)$:
\begin{equation}
A^{(c)} =\frac{1}{2}\,\left(F_{n}^{(c)}(-\om)+F_{n}^{(c)}(\om)\right)  \, ,
\label{B-eq:coeff_A}
\end{equation}
\begin{equation}
B^{(c)}=\frac{1}{2}\,\left(F_{n}^{(c)}(-\om)-F_{n}^{(c)}(\om)\right)\, ,
\label{B-eq:coeff_B}
\end{equation}
\begin{equation}
C^{(c)}=F_{n}^{(c)}(0)\, ,
\label{B-eq:coeff_C}
\end{equation}
and by the functions $K_{n}^{(c)}(\om)$ defined in Eq.~\eqref{eq:Hilberttransf}:
\begin{equation}
U^{(c)} =-\frac{1}{2}\,\left( K_{n}^{(c)}(-\om)+ K_{n}^{(c)}(\om)\right)\, ,
\label{B-eq:coeff_U}
\end{equation}
\begin{equation}
V^{(c)} =-\frac{1}{2}\,\left( K_{n}^{(c)}(-\om)- K_{n}^{(c)}(\om)\right)\, ,
\label{B-eq:coeff_V}
\end{equation}
\begin{equation}
W^{(c)}= K_{n}^{(c)}(0)\, .
\label{B-eq:coeff_W}
\end{equation}

Expanding the products using
Eq.~\eqref{eq:appendix-geometric-projectors} and the definitions in
Eqs.~\eqref{B-eq:coeff_A}--\eqref{B-eq:coeff_W} gives
\begin{equation}
\begin{split}
\Lambda_{ab}=\sum_{c=1}^{N}\Bigl[&
W^{(c)}P_a\chi^{(c)}P_b
-\frac{U^{(c)}}2
\bigl(Q_a\chi^{(c)}Q_b-\Omega_a\chi^{(c)}\Omega_b\bigr)\\
&+\frac{iV^{(c)}}2
\bigl(\Omega_a\chi^{(c)}Q_b+Q_a\chi^{(c)}\Omega_b\bigr)
\Bigr],
\end{split}
\label{B-eq:coherent-blocks}
\end{equation}
\begin{equation}
\begin{split}
\Gamma_{ab}=\sum_{c=1}^{N}\Bigl[&
C^{(c)}P_a\chi^{(c)}P_b
+\frac{A^{(c)}}2
\bigl(Q_a\chi^{(c)}Q_b-\Omega_a\chi^{(c)}\Omega_b\bigr)\\
&-\frac{iB^{(c)}}2
\bigl(\Omega_a\chi^{(c)}Q_b+Q_a\chi^{(c)}\Omega_b\bigr)
\Bigr],
\end{split}
\label{B-eq:dissipative-blocks}
\end{equation}
with $\Lambda_{ba}=\Lambda_{ab}^{\dagger}$ and
$\Gamma_{ba}=\Gamma_{ab}^{\dagger}$.

In particular, the Lamb-shift Hamiltonian is
\begin{equation}
H_L=\frac12\sum_{a,b=1}^{2}\sum_{i,j=1}^{3}
(\Lambda_{ab})_{ij}\sigma_i^{(a)}\sigma_j^{(b)}.
\label{eq:appendix-lamb-blocks}
\end{equation}
For $a\ne b$, operators belonging to different detectors commute;
combining the two ordered contributions and using
$\Lambda_{21}=\Lambda_{12}^{\dagger}$ gives Eq.~\eqref{eq:H_12}.
For $a=b$, the Pauli identity
\begin{equation}
\sigma_i^{(a)}\sigma_j^{(a)}
=\delta_{ij}I_4+i\sum_{k=1}^{3}\varepsilon_{ijk}\sigma_k^{(a)}
\end{equation}
gives the local Lamb-shift contribution up to the additive scalar
$\frac12\sum_a\operatorname{Tr}(\Lambda_{aa})I_4$.
Adding the free detector Hamiltonian gives Eq.~\eqref{eq:H_a}.

For completeness, the fully expanded Cartesian expressions are
listed below. No equality of detector orientations is assumed.
The local Hamiltonian contribution associated with each detector is
\begin{widetext}

\begin{equation}
H_{\mathrm{eff}}^{(a)}
=
\sum_{i=1}^{3}
\bigg\{
\frac{\om}{2}n_i^{(a)}
+
\sum_{c=1}^{N}\sum_{r,s=1}^{3}
\bigg[
\chi_{rs}^{(S,c)}
\frac{V^{(c)}}{2}
\Big(
n_r^{(a)}n_s^{(a)}-\delta_{rs}
\Big)n_i^{(a)}+
\chi_{rs}^{(A,c)}
\frac{U^{(c)}}{2}
\Big(
\varepsilon_{rsi}
+
2\sum_{j=1}^{3}
n_s^{(a)}n_j^{(a)}
\varepsilon_{ijr}
\Big)
\bigg]
\bigg\}\,
\sigma_i^{(a)}
\,.
\label{eq:expanded-24}
\end{equation}
The coherent cross-detector contribution reads
\begin{equation}
\begin{split}
H_{\mathrm{eff}}^{(12)}
&=
\sum_{c=1}^{N}
\sum_{r,s=1}^{3}
\sum_{i,j=1}^{3}
\Bigg[
\chi_{rs}^{(S,c)}
\bigg\{
W^{(c)}
n_r^{(1)}n_s^{(2)}
n_i^{(1)}n_j^{(2)}
-
\frac{U^{(c)}}{2}
\bigg(
\delta_{rs}
\Big(
\delta_{ij}
\sum_{p=1}^{3}
n_p^{(1)}n_p^{(2)}
-
n_j^{(1)}n_i^{(2)}
\Big)
\\
&+
n_r^{(1)}n_s^{(2)}
\Big(
n_i^{(1)}n_j^{(2)}
-
\delta_{ij}
\Big)
+
\delta_{rj}\delta_{si}
\Big(
1-
\sum_{p=1}^{3}
n_p^{(1)}n_p^{(2)}
\Big)+
\delta_{rj}n_s^{(1)}
\Big(
n_i^{(2)}-n_i^{(1)}
\Big)
+
\delta_{si}n_r^{(2)}
\Big(
n_j^{(1)}-n_j^{(2)}
\Big)
\bigg)
\bigg\}
\\
&+
\chi_{rs}^{(A,c)}
\frac{V^{(c)}}{2}
\bigg\{
\sum_{p=1}^{3}
n_p^{(1)}
\varepsilon_{rip}
\Big(
\delta_{sj}
-
n_s^{(2)}n_j^{(2)}
\Big)+
\sum_{p=1}^{3}
n_p^{(2)}
\varepsilon_{rjp}
\Big(
\delta_{si}
-
n_s^{(1)}n_i^{(1)}
\Big)
\bigg\}
\Bigg]
\sigma_i^{(1)}\sigma_j^{(2)}
\,.
\end{split}
\label{eq:expanded-25}
\end{equation}

The dissipative contribution is given by
\begin{equation}
\begin{split}
\mathcal{L}[\rho]
&=
\sum_{c=1}^{N}\sum_{r,s=1}^{3}\sum_{i,j=1}^{3}
\Bigg\{
\chi_{rs}^{(S,c)}
\Bigg[
\sum_{a=1}^{2}
\Bigg(
C^{(c)}n_r^{(a)}n_s^{(a)}n_i^{(a)}n_j^{(a)}
+\frac{A^{(c)}}{2}
\Big(\delta_{rs}-n_r^{(a)}n_s^{(a)}\Big)
\Big(\delta_{ij}-n_i^{(a)}n_j^{(a)}\Big)
\\
&+
\frac{iB^{(c)}}{2}\sum_{p=1}^{3}n_p^{(a)}
\Big[
\varepsilon_{rip}\Big(\delta_{sj}-n_s^{(a)}n_j^{(a)}\Big)
-\varepsilon_{rjp}\Big(\delta_{si}-n_s^{(a)}n_i^{(a)}\Big)
\Big]
\Bigg)
\Bigg(
\sigma_j^{(a)}\rho\sigma_i^{(a)}
-\frac{1}{2}\Big\{\sigma_i^{(a)}\sigma_j^{(a)},\rho\Big\}
\Bigg)
\\
&+
\Bigg(
C^{(c)}n_r^{(1)}n_s^{(2)}n_i^{(1)}n_j^{(2)}
+\frac{A^{(c)}}{2}
\Bigg[
\delta_{rs}
\Bigg(
\delta_{ij}\sum_{p=1}^{3}n_p^{(1)}n_p^{(2)}
-n_j^{(1)}n_i^{(2)}
\Bigg)
+n_r^{(1)}n_s^{(2)}
\Big(n_i^{(1)}n_j^{(2)}-\delta_{ij}\Big)
\\
&+
\delta_{rj}\delta_{si}
\Bigg(1-\sum_{p=1}^{3}n_p^{(1)}n_p^{(2)}\Bigg)
+\delta_{rj}n_s^{(1)}\Big(n_i^{(2)}-n_i^{(1)}\Big)
+\delta_{si}n_r^{(2)}\Big(n_j^{(1)}-n_j^{(2)}\Big)
\Bigg]
\\
&+
\frac{iB^{(c)}}{2}
\Bigg[
\sum_{p=1}^{3}n_p^{(1)}
\varepsilon_{rip}\Big(\delta_{sj}-n_s^{(2)}n_j^{(2)}\Big)
-\sum_{p=1}^{3}n_p^{(2)}
\varepsilon_{rjp}\Big(\delta_{si}-n_s^{(1)}n_i^{(1)}\Big)
\Bigg]
\Bigg)
\\
&\Bigg(
\sigma_j^{(2)}\rho\sigma_i^{(1)}
-\frac{1}{2}\Big\{\sigma_i^{(1)}\sigma_j^{(2)},\rho\Big\}
\Bigg)
+[(1)\leftrightarrow(2)]
\Bigg]
+\chi_{rs}^{(A,c)}
\Bigg[
\sum_{a=1}^{2}
\\
&\Bigg(iA^{(c)}
\Big(
\delta_{ri}\delta_{sj}
+\delta_{rj}n_s^{(a)}n_i^{(a)}
+\delta_{si}n_r^{(a)}n_j^{(a)}
\Big)
-\frac{B^{(c)}}{2}\sum_{p=1}^{3}n_p^{(a)}
\Big[
\varepsilon_{rip}\Big(\delta_{sj}-n_s^{(a)}n_j^{(a)}\Big)
+\varepsilon_{rjp}\Big(\delta_{si}-n_s^{(a)}n_i^{(a)}\Big)
\Big]\Bigg)
\\
&
\Bigg(
\sigma_j^{(a)}\rho\sigma_i^{(a)}
-\frac{1}{2}\Big\{\sigma_i^{(a)}\sigma_j^{(a)},\rho\Big\}
\Bigg)
+\Bigg(
iC^{(c)}n_r^{(1)}n_s^{(2)}n_i^{(1)}n_j^{(2)}
+\frac{iA^{(c)}}{2}
\Bigg[
n_r^{(1)}n_s^{(2)}
\Big(n_i^{(1)}n_j^{(2)}+\delta_{ij}\Big)
\\
&-
\delta_{rj}\delta_{si}
\Bigg(1+\sum_{p=1}^{3}n_p^{(1)}n_p^{(2)}\Bigg)
+\delta_{rj}n_s^{(1)}\Big(n_i^{(2)}+n_i^{(1)}\Big)
+\delta_{si}n_r^{(2)}\Big(n_j^{(1)}+n_j^{(2)}\Big)
\Bigg]
\\
&-
\frac{B^{(c)}}{2}
\Bigg[
\sum_{p=1}^{3}n_p^{(1)}
\varepsilon_{rip}\Big(\delta_{sj}-n_s^{(2)}n_j^{(2)}\Big)
+\sum_{p=1}^{3}n_p^{(2)}
\varepsilon_{rjp}\Big(\delta_{si}-n_s^{(1)}n_i^{(1)}\Big)
\Bigg]
\Bigg)
\\
&\Bigg(
\sigma_j^{(2)}\rho\sigma_i^{(1)}
-\frac{1}{2}\Big\{\sigma_i^{(1)}\sigma_j^{(2)},\rho\Big\}
\Bigg)
+[(1)\leftrightarrow(2)]
\Bigg]
\Bigg\}
\,.
\end{split}
\label{eq:expanded-26}
\end{equation}

\end{widetext}

Here, $\left[(1)\leftrightarrow(2)\right]$ denotes the contribution obtained by interchanging detector labels $(1)$ and $(2)$ in the immediately preceding term.

\section{Scalar-field response functions}
\label{app:response-functions}
This appendix gives explicit response functions for canonically normalized free massless scalar fields along a uniformly accelerated trajectory in integer spacetime dimension $n\geq3$.
We use the Fourier convention of Eq.~\eqref{eq:responsefunc}, with $\mathrm{e}^{-i\omega\tau}$.
Thus a positive frequency $\omega=\om$ corresponds to detector excitation, while $\omega=-\om$ corresponds to decay.
All massless species have the same response, so we omit the index $(c)$ here.
KMS detailed balance also applies to massive fields, but the explicit formulas below assume zero mass.
For proper acceleration $a>0$, the massless Wightman function along the trajectory is~\cite{Takagi86,Davies82}
\begin{equation}
\begin{split}
&\langle0_M|\phi(x(\tau))\phi(x(0))|0_M\rangle\\
&=\frac{\Gamma\left(\frac{n-2}{2}\right)}{4\pi^{n/2}}
\lim_{\epsilon\to0^+}
\left[
\frac{a}{2i\sinh\left(\frac{a}{2}(\tau-i\epsilon)\right)}
\right]^{n-2}.
\end{split}
\label{eq:massless-worldline-wightman}
\end{equation}
The limit is understood in the distributional sense.
Its Fourier transform can be evaluated by contour integration.
For $\omega\neq0$, the result can be written as~\cite{Takagi86,Terashima99}
\begin{equation}
F_n(\omega)
=\frac{\pi}{\omega}
\frac{D_n^M(\omega)d_n(\omega)}
{\mathrm{e}^{\beta\omega}-(-1)^n},
\label{eq:massless-response-n}
\end{equation}
where $\beta=2\pi/a$, the Minkowski density-of-states factor is
\begin{equation}
D_n^M(\omega)
=\frac{2^{2-n}\pi^{(1-n)/2}|\omega|^{n-2}}
{\Gamma\left(\frac{n-1}{2}\right)},
\label{eq:minkowski-density-of-states}
\end{equation}
and
\begin{equation}
d_n(\omega)
=\operatorname{sgn}(\omega)(n-3)!
\left(\frac{ia}{2|\omega|}\right)^{n-3}
R_n(2\omega/a).
\label{eq:dimension-dependent-response-factor}
\end{equation}
The dimensionless polynomial $R_n$ is obtained from the residue at the origin:
\begin{equation}
\begin{split}
R_n(z)&=\operatorname*{Res}_{u=0}
\frac{\mathrm{e}^{-izu}}{(\sinh u)^{n-2}}\\
&=\frac{1}{(n-3)!}
\left.
\frac{d^{n-3}}{du^{n-3}}
\left[
\mathrm{e}^{-izu}
\left(\frac{u}{\sinh u}\right)^{n-2}
\right]
\right|_{u=0}.
\end{split}
\label{eq:residue-polynomial}
\end{equation}
For example,
\begin{equation}
\begin{split}
R_3(z)&=1,\\
R_4(z)&=-iz,\\
R_5(z)&=-\frac{z^2+1}{2}.
\end{split}
\label{eq:residue-polynomial-examples}
\end{equation}
For odd $n$, $d_n(\omega)$ is odd in $\omega$; for even $n$, it is even.
This parity, together with the denominator in Eq.~\eqref{eq:massless-response-n}, reproduces KMS detailed balance.
The Bose-Einstein-like denominator for even $n$ and the Fermi-Dirac-like denominator for odd $n$ constitute the apparent statistics inversion~\cite{Takagi86,Terashima99}.
This is a property of the response spectrum: the field remains bosonic,
and its response satisfies the KMS detailed-balance
relation~\eqref{eq:thermal-detailed-balance}.
For $n=4$, Eqs.~\eqref{eq:massless-response-n}--\eqref{eq:residue-polynomial-examples} give
\begin{equation}
F_4(\omega)
=\frac{\omega}{2\pi\left(\mathrm{e}^{\beta\omega}-1\right)}
\,,
\label{eq:massless-response-four-dimensional}
\end{equation}
where the value at $\omega=0$ is defined by continuity,
$F_4(0)=1/(2\pi\beta)$.
For comparison, in three and five spacetime dimensions one obtains
\begin{equation}
\begin{split}
F_3(\omega)
&=\frac{1}{2\left(\mathrm{e}^{\beta\omega}+1\right)},\\
F_5(\omega)
&=\frac{\omega^2+a^2/4}
{8\pi\left(\mathrm{e}^{\beta\omega}+1\right)}.
\end{split}
\label{eq:massless-response-odd-examples}
\end{equation}
The zero-frequency response is obtained by taking the continuous limit of the complete expression for $F_n(\omega)$, rather than of its individual factors.
In particular,
\begin{equation}
\begin{split}
F_3(0)&=\frac14,\\
F_5(0)&=\frac{a^2}{64\pi}.
\end{split}
\label{eq:massless-response-zero-frequency}
\end{equation}
For the isotropic and mixed couplings considered in
Secs.~\ref{sec:symmetric-coupling}
and~\ref{sec:mixed-coupling}, respectively, the
four-dimensional zero-frequency response equals
the dephasing coefficient $\gamma_0=F_4(0)$.
The massless case $n=2$ is not covered by these formulas and requires a separate infrared prescription.

\section{Inertial subtraction of the Lamb shift}
\label{app:lamb_shift}
For massless scalar fields in four spacetime dimensions, recalling the Hilbert transform of the response function, we have
\begin{equation}
\begin{split}
K_4(\Om)
&=\frac{1}{\pi}\mathcal{P}\int_{-\infty}^{\infty}\mathrm{d}z\,
\frac{F_4(z)}{z-\Om}\\
&=\frac{1}{\pi}\mathcal{P}\int_0^\infty \mathrm{d}z\,
\Big[\frac{F_4(z)}{z-\Om}-\frac{F_4(-z)}{z+\Om}\Big]\\
&=K_4^{\mathrm{(acc)}}(\Om)+K_4^{\mathrm{(vac)}}(\Om)
\, ,
\end{split}
\label{A1}
\end{equation}
where
\begin{equation}
K_4^{\mathrm{(acc)}}(\Om)
=\frac{1}{2\pi^2}\mathcal{P}\int_0^\infty \mathrm{d}z\,
\frac{z}{\mathrm{e}^{\beta z}-1}
\Big(\frac{1}{z-\Om}-\frac{1}{z+\Om}\Big)
\, ,
\label{A2}
\end{equation}
and
\begin{equation}
K_4^{\mathrm{(vac)}}(\Om)
=-\frac{1}{2\pi^2}\mathcal{P}\int_0^\infty \mathrm{d}z\,
\frac{z}{z+\Om}
\, .
\label{A3}
\end{equation}
The inertial vacuum contribution requires ultraviolet
regularization~\cite{Beth47,Milo94,Audre95,Benatti04}.
In this appendix we impose an inertial-reference subtraction
on the full Lamb-shift operator: the contribution generated
by \(K_4^{\mathrm{(vac)}}\) is subtracted from both the
single-detector terms and the induced two-detector interaction.
This prescription defines the renormalized reference model;
it is not, in general, equivalent to a renormalization of the
individual detector gaps alone.
The general generator in the main text retains arbitrary real,
regularized dispersive coefficients, whereas the specialization
used here keeps only \(K_4^{\mathrm{(acc)}}\).
The stationary states derived in the main text are unchanged
by the allowed finite Lamb-shift coefficients because their
commutator with the effective Hamiltonian vanishes.
Transient coherences can depend on these coefficients.
The trajectories starting from \(I_4/4\) considered in Sec.~\ref{sec:unequal-orientations} are an exception, since in the rotated frame~\eqref{eq:rotated-frame} they remain diagonal in the singlet-triplet basis, in which the effective Hamiltonian~\eqref{eq:misaligned-hamiltonian} is also diagonal.

The retained object is odd in its argument, such that
\begin{equation}
\begin{split}
K_4^{\mathrm{(acc)}}(-\Om)&=-K_4^{\mathrm{(acc)}}(\Om),\\
K_4^{\mathrm{(acc)}}(0)&=0
\, .
\end{split}
\label{A4}
\end{equation}

\section{decay of the antisymmetric sector: rank theorem}
\label{app:decay_antisymm}
We show that the antisymmetric components of the density matrix decay exponentially whenever the Kossakowski matrix satisfies $\operatorname{rank}(\Gamma)\geq2$. We assume equal detector orientations and identical couplings to the common environment. Using the collective operators $S_i=\sigma_i^{(1)}+\sigma_i^{(2)}$, the master equation reads
\begin{equation}
\dot{\rho}
=-i[H_{\mathrm{eff}},\rho]
+\sum_{i,j=1}^{3}\Gamma_{ij}
\Big(S_j\rho S_i-\frac{1}{2}\{S_iS_j,\rho\}\Big)
\, .
\label{B1}
\end{equation}
We use the swap operator $U_{\mathrm{sw}}$, defined by
\begin{equation}
U_{\mathrm{sw}}\bigl(|u\rangle\otimes|v\rangle\bigr)
=|v\rangle\otimes|u\rangle
\,.
\end{equation}
The projections onto the singlet and triplet subspaces are
\begin{equation}
\begin{split}
\Pi_s
&=|s\rangle\langle s|
=\frac{I_4-U_{\mathrm{sw}}}{2}\,,\\
\Pi_T
&=I_4-\Pi_s
=\frac{I_4+U_{\mathrm{sw}}}{2}
\,.
\end{split}
\end{equation}
The component of the density operator that is odd under exchange is
\begin{equation}
\begin{split}
\rho_{\mathrm A}
&=\frac{\rho-U_{\mathrm{sw}}\rho U_{\mathrm{sw}}}{2}\\
&=\Pi_T\rho\Pi_s+\Pi_s\rho\Pi_T
\,.
\end{split}
\label{eq:exchange-antisymmetric-component}
\end{equation}
Writing
\begin{equation}
|v\rangle=\Pi_T\rho|s\rangle,
\end{equation}
we have
\begin{equation}
\rho_{\mathrm A}
=|v\rangle\langle s|+|s\rangle\langle v|.
\label{B3}
\end{equation}
Therefore, decay of \(|v\rangle\) implies decay of the entire
exchange-antisymmetric operator sector.
The collective operators annihilate the singlet, $S_i|s\rangle=0$. Moreover, the effective Hamiltonian commutes with the exchange operator and therefore preserves the singlet and triplet subspaces. We write $H_{\mathrm{eff}}|s\rangle=E_s|s\rangle$ and denote its restriction to the triplet subspace by $H_T$. Taking the singlet-triplet block of Eq.~\eqref{B1}, we obtain
\begin{equation}
\frac{d}{d\tau}|v\rangle
=\Big[-i(H_T-E_sI_T)-\frac{1}{2}D_T\Big]|v\rangle
\, ,
\label{B4}
\end{equation}
where $I_T$ is the identity on the triplet subspace $\mathcal H_T=\operatorname{Ran}\Pi_T$, and $D_T$ denotes the restriction to this subspace:
\begin{equation}
D_T=\sum_{i,j=1}^{3}\Gamma_{ij}(S_iS_j) \Big|_{\mathcal H_T}
\, .
\label{B5}
\end{equation}
To evaluate $D_T$, we use the orthonormal triplet basis $|t_k\rangle=\sigma_k^{(1)}|s\rangle$. The collective operators act on this basis as
\begin{equation}
S_i|t_j\rangle
=2i\sum_{k=1}^{3}\varepsilon_{ijk}|t_k\rangle
\, .
\label{B6}
\end{equation}
Consequently,
\begin{equation}
\begin{split}
(D_T)_{ab}
&=\sum_{i,j=1}^{3}\Gamma_{ij}
\langle t_a|S_iS_j|t_b\rangle\\
&=4\sum_{i,j=1}^{3}\Gamma_{ij}
\Big(\delta_{ij}\delta_{ab}-\delta_{ib}\delta_{aj}\Big)\\
&=4\Big[\operatorname{Tr}(\Gamma)\delta_{ab}-\Gamma_{ba}\Big]
\,.
\end{split}
\label{B7}
\end{equation}
Thus, in this basis,
\begin{equation}
D_T=4\Big[\operatorname{Tr}(\Gamma)I_T-\Gamma^T\Big]
\, .
\label{B8}
\end{equation}
Since $\Gamma$ is Hermitian and positive semidefinite, its eigenvalues $\gamma_1,\gamma_2,\gamma_3$ are nonnegative. The transpose has the same eigenvalues, so the eigenvalues of $D_T$ are
\begin{equation}
\begin{split}
d_1&=4(\gamma_2+\gamma_3),\\
d_2&=4(\gamma_1+\gamma_3),\\
d_3&=4(\gamma_1+\gamma_2)
\, .
\end{split}
\label{B9}
\end{equation}
All three are strictly positive if and only if at least two eigenvalues of $\Gamma$ are strictly positive. Hence, 
\begin{equation}
D_T>0
\Longleftrightarrow
\operatorname{rank}(\Gamma)\geq2
\, .
\label{B10}
\end{equation}
Let $d_{\min}>0$ denote the smallest eigenvalue of $D_T$. Since $H_T-E_sI_T$ is Hermitian, the Hamiltonian term does not contribute to the derivative of the squared norm in Eq.~\eqref{B4}. Therefore,
\begin{equation}
\begin{split}
\frac{d}{d\tau}\|v(\tau)\|^2
&=-\langle v(\tau)|D_T|v(\tau)\rangle\\
&\leq-d_{\min}\|v(\tau)\|^2
\, ,
\end{split}
\label{B11}
\end{equation}
which implies
\begin{equation}
\|v(\tau)\|
\leq\mathrm{e}^{-d_{\min}\tau/2}\|v(0)\|
\, .
\label{B12}
\end{equation}
By Eq.~\eqref{B3}, all antisymmetric components consequently decay exponentially to zero. Thus $\operatorname{rank}(\Gamma)\geq2$ is a sufficient condition for the asymptotic decay of the complete antisymmetric sector, including the Hamiltonian contribution.

\bibliography{Bibliography.bib}

\end{document}